\documentclass[
prd,
lengthcheck,
floats,
floatfix,
nofootinbib,
tightenlines,
showpacs]{revtex4}
\usepackage{amsmath}
\usepackage{amssymb}
\usepackage{graphicx}
\usepackage[usenames]{color}

\usepackage{ulem}
\newcommand{\half}{{\textstyle \frac{1}{2}}}
\newcommand{\lamzer}{\lambda_0}
\newcommand{\lamzerp}{\lambda_0{\hspace{-1.5mm}}'}
\newcommand{\lamone}{\lambda_1}
\newcommand{\lamonep}{\lambda_1{\hspace{-1.5mm}}'}
\newcommand{\lamtwo}{\lambda_2}
\newcommand{\lamtwop}{\lambda_2{\hspace{-1.5mm}}'}
\newcommand{\laminf}{\lambda_\infty}
\newcommand{\laminfp}{\lambda_\infty{\hspace{-3.0mm}}'{\hspace{3.0mm}}}
\newcommand{\shift}{\beta}
\newcommand{\ljump}{\big[\!\big[}
\newcommand{\rjump}{\big]\!\big]}
\begin{document}

\title{Persistent junk solutions in time-domain\\
       modeling of extreme mass ratio binaries}
      \author{
              Scott E.~Field${}^{1,}$\footnote{
              {\tt Scott\_Field@brown.edu},
              ${}^\dagger\,{}${\tt Jan\_Hesthaven@brown.edu},
              ${}^\ddagger\,{}${\tt srlau@math.unm.edu}},
              Jan~S.~Hesthaven${}^{2,\dagger}$, and
              Stephen R.~Lau${}^{3,\ddagger}$
              }
     \affiliation{
              ${}^1$Department of Physics, Brown
               University, Providence, RI 02912\\
              ${}^2$Division of Applied Mathematics, Brown
               University, Providence, RI 02912\\
              ${}^3$Mathematics and Statistics, University
               of New Mexico, Albuquerque, NM 87131
             }

%%%%%%%%%%%%%%%%%%%%%%%%%%%%%%%%%%%%%%%%%%%%%%%%%%%%%%%%%%%%%%%
\begin{abstract}
In the context of metric perturbation theory for non-spinning 
black holes, extreme mass ratio binary (EMRB) systems are 
described by distributionally forced master wave equations. 
Numerical solution of a master wave equation as an initial 
boundary value problem requires initial data. However, because 
the correct initial data for generic-orbit systems is unknown, 
specification of trivial initial data is a common choice, 
despite being inconsistent and resulting in a solution which 
is initially discontinuous in time. As is well known, this 
choice leads to a ``burst" of junk radiation which eventually 
propagates off the computational domain. We observe another 
potential consequence of trivial initial data: development 
of a persistent spurious solution, here referred to as the 
{\em Jost junk solution}, which contaminates the physical 
solution for long times. This work studies the influence of 
both types of junk on metric perturbations, waveforms, and 
self-force measurements, and it demonstrates that smooth 
modified source terms mollify the Jost solution and reduce 
junk radiation. Our concluding section discusses the 
applicability of these observations to other numerical
schemes and techniques used to solve distributionally 
forced master wave equations.
\end{abstract}
\maketitle
%%%%%%%%%%%%%%%%%%%%%%%%%%%%%%%%%%%%%%%%%%%%%%%%%%%%%%%%%%%%%%%
%
%  INTRODUCTION
%
%%%%%%%%%%%%%%%%%%%%%%%%%%%%%%%%%%%%%%%%%%%%%%%%%%%%%%%%%%%%%%%
\section{Introduction}\label{sec:intro}
Extreme mass ratio binary (EMRB) systems are typically comprised of 
a small compact object, such as a stellar black hole, orbiting a 
super-massive blackhole, and the gravitational radiation generated
by such systems is potentially detectable by the LISA project. 
A number of approaches attempt to model the resulting gravitational 
waveforms, including effective one body formulations 
\cite{Damour,Pan,NDT2007}, 
effective field theory techniques \cite{GalleyEFT,Galley}, 
post-Newtonian expansions \cite{BlanchetPN}, 
self-force effects 
\cite{Detweiler2009Rev,Detweiler2005Rev,Barack2009Rev,Tanaka2005Rev}, 
and different gauge choices \cite{BarackSago,BMNOS2002,SagoCapra}.
When including high-order effects or performing comparisons between 
techniques, improved EMRB modeling will increasingly require the
identification and reduction of all error sources (both numerical 
and systematic).

Consider a small perturbation $h_{\mu\nu}$ of a fixed background 
Schwarzschild metric, where $h_{\mu\nu}$ 
satisfies the linearized Einstein equations. The metric perturbation 
$h_{\mu\nu}$ describing an EMRB can be reconstructed from a collection 
of scalar {\em master functions} $\Psi$, each of which obeys a forced 
wave equation of the form (with all multipole indices suppressed)
\begin{align}\label{eq:genericwaveeq}
\begin{split}
& -\partial_t^2 \Psi + \partial_x^2 \Psi 
  -V(r) \Psi 
\\
& = f(r) \big[
G(t,r) \delta(r-r_p(t)) + F(t,r) \delta'(r-r_p(t))\big].
\end{split}
\end{align}
The coordinates here are the areal radius $r$, the Regge--Wheeler 
tortoise coordinate $x = r + 2M\log(\frac{1}{2}r/M - 1)$, and the 
time-dependent radial location $r_p(t)$ of the smaller mass or 
``particle". $M$ is the mass parameter of the background solution, 
$f(r) = 1-2M/r$, and $V(r)$ is either the Regge-Wheeler or Zerilli 
potential (explicit expressions for both are given in 
Sec.~\ref{sec:StaticMasterEqn}). The 
{\em distributional} inhomogeneity on the right--hand side of 
\eqref{eq:genericwaveeq} involves Dirac delta functions, as well 
as the ordinary functions $F(t,r)$ and $G(t, r)$. For all possible 
choices of the master function, $F(t,r)$ and $G(t, r)$ are listed in, 
for example, in Refs.~\cite{Martel_GravWave,dG_EMRB}. Here it suffices 
to note that their evaluation requires knowledge of the particle's 
four-velocity $u^{\alpha}$, orbital energy and angular momentum 
parameters ($E_p$, $L_p$), and equatorial location 
$(r_p(t),\pi/2,\phi_p(t))$. In the model we study, integration of 
the geodesic equations determines the timelike particle trajectory 
$(r_p(t),\phi_p(t))$ in the equatorial plane $\theta = \pi/2$. 
\cite{Chan_MathBlack,SopuertaLaguna,CUT,Martel_GravWave}

One approach for computing EMRB waveforms is to numerically solve 
Eq.~(\ref{eq:genericwaveeq}) as a time-domain initial value problem
with prescribed initial data. The exact initial data for generic
point-particle trajectories is non-trivial, and the most 
common choice is therefore to set both $\Psi$ and its time 
derivative to zero. 
(See Refs.~\cite{MartelRadialIC,
PriceLateTimeTails,
Campanelli1,
Campanelli2}
for the construction of more realistic data.) Inspection of 
(\ref{eq:genericwaveeq}) shows that trivial data is inconsistent with 
the jump conditions stemming from the delta function terms in the 
inhomogeneity. As a result, trivial data results in an impulsive 
(i.~e.~discontinuous in time) start-up. This paper addresses the main 
question of if, and when, a {\em physical} solution eventually emerges 
from such trivial initial data. Ideally, we would have both the correct 
source terms and initial conditions. Without the exact initial data, we 
consider modifying the source terms such that they are consistent with 
the choice of trivial initial data. Precisely, the source terms 
are ``switched on" smoothly via the following prescription:
\begin{align}\label{eq:smoothFandG}
\begin{split}
& F(t,r) \rightarrow F(t,r) \times
\\
& \left\{\begin{array}{rcl}
{\textstyle \frac{1}{2}}
[\mathrm{erf}(\sqrt{\delta}(t - t_0 - \tau/2)+1]
& & \text{for } t_0\leq t \leq t_0+\tau\\
1 & & \text{for } t > t_0+\tau,
\end{array}\right.
\end{split}
\end{align}
and the same for $G(t,r)$. Typically, the initial time $t_0
= 0$, and the timescale $\tau$ is much shorter than the final time
of the run. Choosing suitable $\tau$ and $\delta$, one achieves
smooth and consistent start-up to machine precision.

To appreciate some of the issues associated with the main question
above, consider a particle in a fixed circular orbit. The energy 
$\dot{E}_{GW}$ and angular momentum $\dot{L}_{GW}$ luminosities 
for gravitational waves are then constant in time and 
obey the relation $\dot{E}_{GW} = \Omega \dot{L}_{GW}$, where 
$\Omega$ is the angular velocity of the particle. However, 
verification of this relationship is limited by a finite 
computational domain, leading to an $O(r^{-1})$ error (see 
Ref.~\cite{Zenginoglu} for a recent suggestion
towards overcoming this limitation). Therefore, 
numerical verification of
$\dot{E}_{GW} = \Omega \dot{L}_{GW}$  
is a useful diagnostic only in the distant wave-zone.
In the near-zone we might also test 
``$\dot{E}_{GW} = \Omega \dot{L}_{GW}$", now constructing the 
luminosities with self-force quantities via 
(\ref{eq:EL_evolve}) below; however, because $\Psi$ is 
discontinuous at the particle location, self-force measurements
will involve large errors unless due care is taken. For generic 
quasi-periodic orbits, selection of a meaningful set of 
diagnostics is not straightforward. In particular, we can 
neither infer steady-state behavior throughout the computational 
domain, nor claim we have a solution which solves the hypothetical 
``true" initial value boundary problem. These difficulties are 
due to the inconsistent initial conditions. That is, we are really 
solving a problem different from the physical one.
As a partial resolution of these issues, we
examine a direct test condition which is necessary to claim that 
a physically correct solution has been achieved everywhere in 
the computational domain. This is a simple 
self-consistency condition relating the Cunningham-Price-Moncrief 
(CPM) and Regge-Wheeler (RW) master functions. Violations of this 
relationship are necessarily due to numerical errors and/or 
incorrect initial conditions.

We will refer to errors seeded by the initial conditions 
as ``junk". One type of junk either propagates off the 
computational domain or decays away. We collectively refer to 
such junk radiation, junk quasi-normal ringing, and junk Price 
tails as {\em dynamical junk}. The key observation of this paper
is that trivial initial conditions {\em may} also give rise to a static 
distributional junk solution $\Psi_{\mathrm{Jost}}$, which we 
refer to as {\em Jost junk}. In terms of the ``Schr\"{o}dinger 
operator" $H = -\partial^2_x + V$, a Jost solution satisfies $H
\Psi_\mathrm{Jost}^\pm = \nu^2 \Psi_\mathrm{Jost}^\pm$,
with $\Psi_\mathrm{Jost}^\pm \sim \exp(\pm\mathrm{i}\nu x)$
as $x\rightarrow\infty$ \cite{Donninger2009}. In this paper, we are
exclusively interested in ``zero-energy" Jost solutions for which
$\nu = 0$, in which case $\Psi_\mathrm{Jost}$ does not behave
exponentially at infinity (see below). Therefore, in what follows a 
Jost function satisfies a ``zero-energy", time-independent, 
Schr\"{o}dinger equation $(-\partial^2_x + V)\Psi_{\mathrm{Jost}}=0$ 
to the left and right of the particle, and, as it turns out, is 
discontinuous at the particle location. We find that 
$\Psi_{\mathrm{Jost}}$ has a non-negligible effect in the wave-zone, 
yet is often small enough to be buried into the $O(r^{-1})$ error 
associated with a waveform ``read-off" in the far-field. 
\begin{table}
\begin{tabular}{||rcl||}
\hline
$a,b$: & & Endpoint of computational domain $[a,b]$.\\
$S_L,S_R$: & & Number of subdomains to left and right of particle.\\
$N$: & & Number of points on each subdomain.\\
$\tau,\delta$: & & Smoothing parameters
               introduced in Eq.~(\ref{eq:smoothFandG}).\\
$\Delta t, t_F$: & & Timestep and final time.\\
$M=1$: & & Schwarzschild mass parameter.\\
$m_p=1$: & & Particle mass.\\
\hline
\end{tabular}
\caption{{\sc Basic set of parameters for a numerical simulation.}
This set is not complete, but in what follows we often refer to 
these variables. For all our simulations $M = 1 = m_p$, 
where the choice $m_p = 1$ is equivalent to working with 
per-particle-mass perturbations $\Psi/m_p$.}
\label{table:parameters}
\end{table}

We will adopt trivial initial conditions throughout, but 
allow for modified ``smoothed" source terms according to 
the aforementioned description (\ref{eq:smoothFandG}). 
Our chief goal is to study the properties of 
the numerical solutions computed with and without smoothed 
source terms, especially in the context of the Jost solution. 
To carry out numerical simulations, we have primarily used the
nodal Legendre discontinuous Galerkin method described 
in Ref.~\cite{dG_EMRB}, and further details of this method 
will not be given here. In addition, some of our results have 
either been obtained or independently verified with a nodal
Chebyshev method (similar to the one described in 
Refs.~\cite{CanizaresSopuerta,CanizaresSopuerta1}), 
which also features multiple subdomains and upwinding. 
Our nodal Chebyshev method treats the jump 
discontinuities at the particle location in the same fashion 
as outlined in Ref.~\cite{dG_EMRB} for the nodal dG method. 
Both our dG and Chebyshev methods solve a first order 
system representing (\ref{eq:genericwaveeq}). Namely,
\begin{subequations}\label{eq:firstordersys}
\begin{align}
\partial_\lambda\Psi & = \shift^\xi \Phi - \Pi \\
\begin{split}
\partial_\lambda\Pi & = \shift^\xi \partial_\xi \Pi
-(\partial x/\partial\xi)^{-1}\partial_\xi 
  [(\partial x/\partial\xi)^{-1}\Phi]
\\
& + V(r) \Psi + J_1 \delta(\xi-\xi_p)
\end{split}
\\
\partial_\lambda \Phi & = \partial_\xi (\shift^\xi \Phi)
- \partial_\xi \Pi  + J_2 \delta(\xi-\xi_p),
\end{align}
\end{subequations}
where the time-space coordinates $(\lambda,\xi)$ are adapted to the 
particle history (the particle location $\xi = \xi_p$ remains fixed
in this system). Eq.~(\ref{eq:firstordersys}a) defines 
$\Pi$, the variable\footnote{In our approach, from all fields we 
explicitly remove delta function terms arising from the distributional
inhomogeneity. Therefore, $\Phi = \partial_\xi\Psi$ does {\em not}
hold in the sense of distributions. More precisely, in the case of 
circular orbits, our $\Phi$ is 
$\partial_x\Psi - \ljump\Psi\rjump\delta(x -x_p)$.}
$\Phi = \partial_\xi\Psi$, and Ref.~\cite{dG_EMRB}
relates the $\lambda$--dependent jump terms $J_{1,2}$ to the
sources in (\ref{eq:genericwaveeq}). Most of this paper considers 
circular orbits, for which $\lambda = t$, $\xi = x$, and the
shift vector $\shift^\xi = 0$. We often refer to the variables
$\Pi$ and $\Phi$ below, and for circular orbits these are
$-\partial_t\Psi$ and $\partial_x\Psi$, respectively.
Throughout the paper, we make reference to the 
parameters listed in Table \ref{table:parameters}.

This paper is organized as follows. Section \ref{sec:JostJunk} focuses
on the Jost solution, from both empirical and analytical standpoints.
Here we present analytic formulas for Jost solutions and compare
them with numerical results. Section \ref{sec:consequences} considers 
several practical consequences of impulsive start-up for EMRB modeling
with circular orbits: violation of the axial consistency condition, 
contamination of waveform luminosities, and influence on 
self-force measurements. This section also gives a preliminary report 
on consequences for eccentric orbits. Concluding remarks are given in 
Sec.~\ref{sec:conclusions}, where we touch upon finite-difference 
methods while discussing the universality of our results.
Longer calculations appear in the Appendix. 

%%%%%%%%%%%%%%%%%%%%%%%%%%%%%%%%%%%%%%%%%%%%%%%%%%%%%%%%%%%%%%%
%
%  SECTION 2: JOST SOLUTION
%
%%%%%%%%%%%%%%%%%%%%%%%%%%%%%%%%%%%%%%%%%%%%%%%%%%%%%%%%%%%%%%%
\section{Jost solution} \label{sec:JostJunk}
To better explain the origin of the Jost junk solution, we first consider
a toy model: the ordinary 1+1 wave equation with distributional forcing.
We then examine the Jost solution for the master wave equations, 
with a forcing determined by a circular orbit. 
%*%*%*%*%*%*%*%*%*%*%*%*%*%*%*%*%*%*%*%*%*%*%*%*%*%*%*%*%*%*%*%
\begin{figure}
\centering
\includegraphics[height=2.00in]{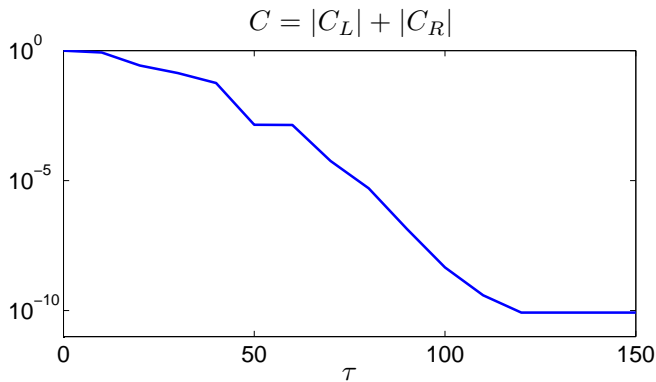}
\caption{{\sc Dependence of $C$ on smoothing parameters.}
We have empirically determined that $|C_L| = \frac{1}{2} = |C_R|$ 
for an impulsive start-up, corresponding to $C = 1$ at the leftmost 
point.  The parameter $\delta$ is different for each $\tau$; 
$\delta = 2$ for $\tau = 10$ and $\delta = 0.0058$ for $\tau = 150$.
}
\label{fig:1P1_constant_smoother}
\end{figure}
%*%*%*%*%*%*%*%*%*%*%*%*%*%*%*%*%*%*%*%*%*%*%*%*%*%*%*%*%*%*%*% 

\subsection{Forced 1+1 wave equation} \label{sec:1p1}
For a fixed velocity $v$ obeying $|v| < 1$, we 
consider the model
\begin{align}
\begin{split}\label{eq:forced1p1}
-\partial_t^2\Psi + \partial_x^2\Psi 
& =                 G(t)\delta(x-vt) 
+                  F(t)\delta'(x-vt)
\\ 
G(t) & = \cos t = -\mathrm{i}F(t).
\end{split}
\end{align}
Ref.~\cite{dG_EMRB} has shown that 
\begin{align}\label{eqn:1p1_sol}
\begin{split}
\Psi(t,x) & = -\half\sin\vartheta 
          +  \half\mathrm{i}\gamma^2 
             [v+\mathrm{sgn}(x-vt)]\cos\vartheta
\\
\vartheta & = \gamma^2(t-xv-|x-vt|)
\end{split}
\end{align}
is an exact particular solution to (\ref{eq:forced1p1}). Here 
$\gamma = (1-v^2)^{-1/2}$ is the usual relativistic factor. For this 
model, junk radiation propagates off the computational domain with 
speeds $\pm 1$. However, when numerically solving this equation 
subject to (incorrect) trivial initial conditions, we observe that 
the numerical solution no longer converges to the particular 
solution. For simulations involving (\ref{eq:forced1p1}), we have 
used the dG method with (cf.~Table \ref{table:parameters})
$a = -100$, $b = 100$, $S_L = 10$,
$S_R = 10$, $N = 27$, and $\Delta t = 0.01$.
To compute errors relative to the exact solution, 
we have first interpolated onto a uniformly spaced $x$--grid with 
5121 points. Furthermore, to better model the circular orbit 
scenario for EMRBs, we have taken $v = 0$. 

With the exact solution used to generate initial conditions 
at $t=0$, the nodal dG method exhibits spectral convergence 
throughout the computational domain (and for all fields with the 
wave equation treated as a first order system) \cite{dG_EMRB}. 
However, with trivial initial conditions, only the 
corresponding numerical derivatives, $\Pi_\mathrm{numerical}$ 
and $\Phi_\mathrm{numerical}$, converge to the correct values, 
whereas $\Psi_\mathrm{numerical}$ itself is off by a constant 
value on each subdomain. Let us write 
\begin{align}
\Psi_{\mathrm{numerical}}
=\left(\Psi_L + C_L\right)\Theta(-x)
+\left(\Psi_R + C_R \right)\Theta(x) ,
\end{align}
where $\Theta(x)$ is the Heaviside function and the exact solution 
from (\ref{eqn:1p1_sol}) is
\begin{align}
\begin{split}
\Psi_L
& = -\half\sin(t+x) - \half\mathrm{i}\cos(t+x)
\\
\Psi_R
& = -\half\sin(t-x) + \half\mathrm{i}\cos(t-x).
\end{split}
\end{align}
We introduce the time-independent 1+1 Jost junk solution
\begin{equation}
\Psi_\mathrm{Jost} = C_L \Theta(-x) + C_R \Theta(x),
\end{equation}
in order to express the numerical solution as 
$\Psi_{\mathrm{numerical}} = \Psi_{\mathrm{exact}} + \Psi_{\mathrm{Jost}}$. 

We examine the dependence of $C=|C_L|+|C_R|$ on the smoothing parameters 
$(\tau, \delta)$, defined analogously to those in (\ref{eq:smoothFandG}),
but here introduced to smooth our toy source term $\cos t\delta(x) + 
\mathrm{i}\cos t\delta'(x)$. We restrict the parameter space by first 
choosing $\tau$, and then finding the smallest $\delta$ such that 
${\textstyle \frac{1}{2}}[\mathrm{erf}(\sqrt{\delta}(t - t_0 - \tau/2)+1
]$ is less than $10^{-16}$ when $t=0$ and greater than $1-10^{-16}$ when 
$t=\tau$. These requirements ensure that the start-up phase is smooth to 
machine precision, while providing the most gradual rate at which the 
distributional source terms are turned-on. 
Figure \ref{fig:1P1_constant_smoother} shows that the troublesome 
constant term is arbitrarily well suppressed by the smoothing procedure.
However, we find that the value of $C$ remains fixed when varying 
the timestep. The final run time for each data point in the plot is 
$t_F =\tau + 150$. No essential difference exists 
between the $v=0$ and $v\neq 0$ cases, except that for the latter case 
we must ensure that the particle does not get too close to the boundary.
Let $\Psi_\mathrm{smooth}$ represent $\Psi_\mathrm{numerical}$ obtained
with smoothing, and $\Psi_\mathrm{impulsive}$ represent 
$\Psi_\mathrm{numerical}$ obtained without smoothing. Then we have
shown $\Psi_\mathrm{smooth} \simeq \Psi_\mathrm{exact}$, so that
\begin{equation}\label{eqn:empiricalJost}
\Psi_\mathrm{Jost} \simeq  \Psi_\mathrm{impulsive} - \Psi_\mathrm{smooth}  
\end{equation}
is another expression for the Jost solution, valid up to method error.
In the next subsection we consider this expression in the context of
master wave equations.

\subsection{Master wave equations}\label{sec:StaticMasterEqn}
\begin{figure}
\centering
\includegraphics[height=2.9in]{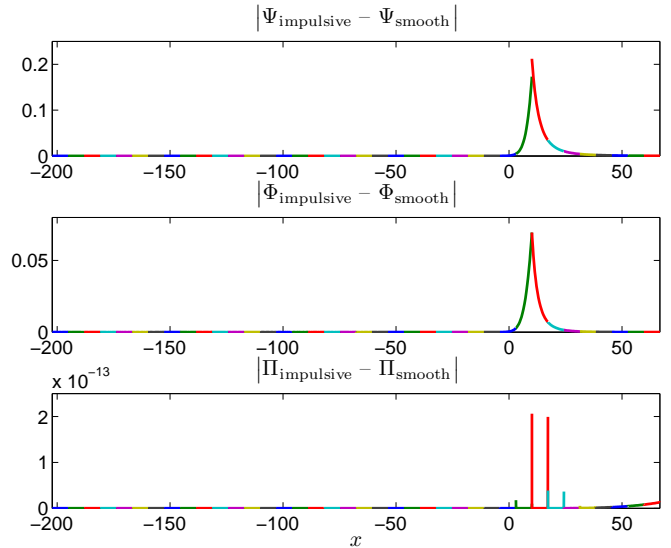}
\caption{
{\sc  Difference between smoothly and impulsively started
      CPM fields.} Here $\ell = 3$, $m = 2$, 
      and the snapshot is taken at $t = 600$. 
}
\label{fig:Smooth_vs_imp_AllFields_circ_cpm}
\end{figure}

The first numerical experiment in this subsection involves the axial 
sector with
\begin{equation}\label{eq:ReggeWheelerV}
V^\mathrm{axial}(r) = \frac{f(r)}{r^2}\left[\ell(\ell+1)
- \frac{6M}{r}\right]
\end{equation}
in (\ref{eq:genericwaveeq}),
and assumes CPM source terms (see the appendix of \cite{dG_EMRB} 
for the precise expressions). To empirically verify that an impulsive
start-up also leads to a Jost solution in this setting, we will form
and plot the expression (\ref{eqn:empiricalJost}), 
using the Chebyshev method. Later on, we will
give analytic expressions for static Jost solutions. The experiment 
enforces Sommerfeld boundary conditions at the left physical boundary, 
and radiation outer boundary conditions \cite{LAU,dG_EMRB} on the right
boundary. Our smoothing parameters are $\tau = 150$ and $\delta = 0.0058$. 
We compute the $(\ell,m)=(3,2)$ metric perturbations for a particle in 
circular orbit initially at $(r,\phi)=(7.9456,0)$. Other parameters 
(cf.~Table \ref{table:parameters}) are $a\simeq -202.16$, 
$b = 60 + 2\log(29)\simeq 66.73$, $S_L = 30$, $S_R = 8$, $N = 26$,
$\Delta t \simeq 0.03$, and $t_F = 600$. 
Figure \ref{fig:Smooth_vs_imp_AllFields_circ_cpm} 
shows the result. The plots suggest that the Jost junk solution 
affects $\Psi^\mathrm{CPM}_\mathrm{impulsive}$ and its spatial 
derivatives. 
%*%*%*%*%*%*%*%*%*%*%*%*%*%*%*%*%*%*%*%*%*%*%*%*%*%*%*%*%*%*%*% 
%*%*%*%*%*%*%*%*%*%*%*%*%*%*%*%*%*%*%*%*%*%*%*%*%*%*%*%*%*%*%*% 

For both axial and polar perturbations generated by 
circular orbits, we now present the analytic form of the Jost 
solution, suppressing throughout the analysis both orbital 
$\ell$ and azimuthal $m$ indices. For circular orbits we have 
observed empirically that the Jost junk solution can be 
written as
\begin{align}
\label{eqn:zeroPsi_general}
\begin{split}
\Psi_{\mathrm{Jost}}^{\mathrm{axial/polar}} & =
\\
& C_Lv_L^{\mathrm{axial/polar}}\Theta(-x)+
C_Rv_R^{\mathrm{axial/polar}}\Theta(x),
\end{split}
\end{align}
where $C_L$ and $C_R$ are complex constants. The functions
$v_{L,R}^{\mathrm{axial/polar}}$ satisfy a 
Schr\"{o}dinger equation $Hv = 0$ defined by the operator
\begin{align}
H^{\mathrm{axial/polar}} = -\partial^2_x
                           +V^{\mathrm{axial/polar}},
\end{align}
where $V^{\mathrm{axial}}$ is given in Eq.~(\ref{eq:ReggeWheelerV})
and, in terms of $n=\frac{1}{2}(\ell-1)(\ell+2)$,
\begin{align}\label{eq:PolarV}
\begin{split}
V^\mathrm{polar}(r) & = \frac{2f(r)}{(n r + 3 M)^2}\times
\\
&
\left[n^2 \left(1 + n + \frac{3M}{r}\right) +
\frac{9M^2}{r^2} \left(n + \frac{M}{r}\right)\right].
\end{split}
\end{align}
The functions
$v_{L}^{\mathrm{axial/polar}}$ satisfy the Schr\"{o}dinger equation
to the left of the particle, and the functions
$v_R^{\mathrm{Axial/Polar}}$ the equation to the right. The relevant 
solutions to $Hv = 0$ decay either as $r \rightarrow 2M^+$ or 
$r \rightarrow \infty$. 

We derive expressions for all four functions 
$v_{L,R}^{\mathrm{axial/polar}}$ in the Appendix, 
adopting the dimensionless radius $\rho = (2M)^{-1}r$ as the basic 
variable. Here we record the set of axial functions,
\begin{subequations}\label{eqn:axialvLvR}
\begin{align}
v^\mathrm{axial}_L(\rho) & =
  \rho^{-\ell} {}_2 F_1(\ell+\jmath+1,\ell-\jmath+1;1;(\rho-1)/\rho)
\\
v^\mathrm{axial}_R(\rho) & =  
  \rho^{-\ell} {}_2 F_1(\ell +\jmath +1,\ell - \jmath +1; 2(\ell +
1);\rho^{-1}),
\end{align}
\end{subequations}
where for gravitational perturbations the spin $\jmath = 2$.
Evidently, up to transformations of the dependent and independent 
variables, the equation $H^\mathrm{axial}v = 0$ is the hypergeometric 
equation. The equation $H^\mathrm{polar}v = 0$ involves an extra regular 
singular point, and its normal form is a particular realization of 
the Heun equation. Nevertheless, by exploiting 
certain intertwining relations between the polar and axial master 
functions \cite{AndersonPrice1991}, we are likewise able to express 
$v^\mathrm{polar}_{L,R}$ in terms of the classical 
Gauss-hypergeometric function ${}_2F_1$. The Appendix 
gives further details.
%*%*%*%*%*%*%*%*%*%*%*%*%*%*%*%*%*%*%*%*%*%*%*%*%*%*%*%*%*%*%*% 
\begin{figure*}
\centering
% trim=l b r t
\includegraphics[clip=true,height=2.5in,trim=0 20mm 0 0]{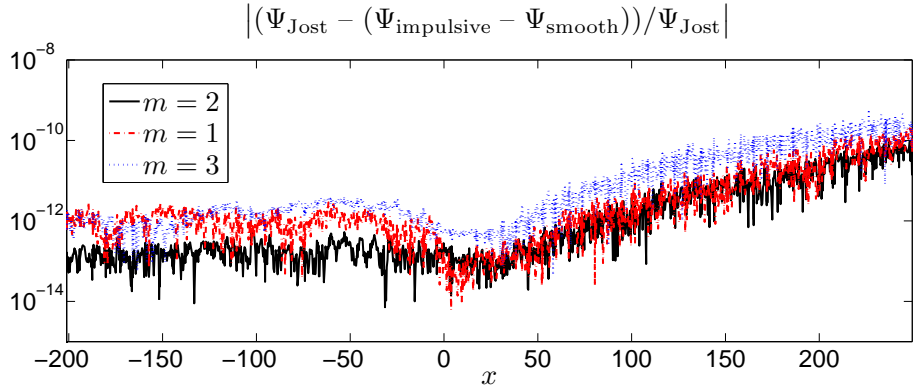}
\caption{{\sc Comparison between analytic and 
              numerical Jost solutions.}
CPM and ZM modes respectively correspond to 
$(\ell,m)=(3,2)$ and $(\ell,m)=(3,1),(3,3)$.}
\label{fig:L3_zeroECompare}
\end{figure*}
%*%*%*%*%*%*%*%*%*%*%*%*%*%*%*%*%*%*%*%*%*%*%*%*%*%*%*%*%*%*%*% 

To complete our analytic expressions for the Jost solutions,
we still must determine $C_L$ and $C_R$. Our notation for a 
time--dependent jump is, for example,
\begin{align}
\ljump\Psi\rjump(t) & \equiv
\lim_{\epsilon \rightarrow 0^+}
\big[
\Psi(t,r_p(t)+\epsilon)-
\Psi(t,r_p(t)-\epsilon)\big] 
\nonumber\\
& =
\lim_{\epsilon \rightarrow 0^+}
\big[
\Psi(t,r_p+\epsilon)-
\Psi(t,r_p-\epsilon)\big],
\label{eq:jumpdefinition}
\end{align}
with the last equality holding for a circular orbit. As derived in
Ref.~\cite{dG_EMRB}, for a circular orbit the analytical jump 
determined by Eq.~(\ref{eq:genericwaveeq}) is
\begin{align} \label{eq:jumpPsiAnal}
\ljump\Psi_\mathrm{analytic}\rjump(t) & = \frac{F(t,r_p)}{f_p},
\end{align}
where the subscript ``p" indicates evaluation at the particle
location. For trivial initial data (that is $\Psi = 0$) this
jump will in general not be satisfied at $t=0$. 
We find empirically that the jump in $\Psi_{\mathrm{Jost}}$ 
exactly cancels $\ljump\Psi_\mathrm{analytic}\rjump(0)$, 
while the jump in $\partial_x \Psi_{\mathrm{Jost}}$ is zero. 
The system of equations used to determine our constants
is therefore
\begin{align} 
\begin{split}
v_R(r_p)C_R-v_L(r_p)C_L & = -\frac{F(0,r_p)}{f_p}
\\
v_R'(r_p)C_R-v_L'(r_p)C_L & = 0,
\end{split}
\label{eqn:constantEqn1}
\end{align}
which has solution
\begin{align} 
\begin{split}
C_R  & = -\frac{F(0,r_p)}{f_p}
\left(\frac{v_L'}{v_R v_L' - v_L v_R'}\right)_p
\\
C_L  & = C_R \left(\frac{v_R'}{v_L'}\right)_p.
\end{split}
\label{eqn:constantsEqn2}
\end{align}
Recall that $\Psi_{\mathrm{Jost}}$ may be numerically approximated
as $\Psi_{\mathrm{impulsive}} - \Psi_{\mathrm{smooth}}$ 
[cf.~Eq.~(\ref{eqn:empiricalJost})].
Figure \ref{fig:L3_zeroECompare} depicts
the relative error 
$\big|(\Psi_{\mathrm{Jost}} - (\Psi_{\mathrm{impulsive}} - 
\Psi_{\mathrm{smooth}}))/\Psi_{\mathrm{Jost}}\big|$ 
for $\ell=3$ perturbations,
with $\Psi_{\mathrm{Jost}}$ given by (\ref{eqn:zeroPsi_general}).
To generate this figure, we have used nearly the same set-up 
as described for Fig.~\ref{fig:Smooth_vs_imp_AllFields_circ_cpm},
but with the outer boundary $b = 240 + 2\log(119)$ and final
time $t_F = 3100$.

\subsection{Jost solution and radiation boundary conditions}
We wish to examine the extent to which the right analytic Jost 
solutions $v^\mathrm{axial/polar}_R$ satisfy radiation 
boundary conditions based on Laplace convolution \cite{AGH,LAU}, 
as these are boundary conditions adopted for our numerical simulations. 
Unfortunately, for blackhole perturbations the issue would seem 
difficult to address analytically. Therefore, we consider the 
analogous issue for the flatspace radial wave equation.

Consider a flatspace multipole solution
$r^{-1}\Psi(t,r)Y_{\ell m}(\theta,\phi)$ to the ordinary 3+1 wave
equation, and assume the multipole is initially of compact support
in radius $r$. Exact non-reflecting boundary conditions relative to
a sufficiently large outer boundary radius $b$ then take the form
\cite{AGH}
\begin{align}\label{eq:nbc}
\begin{split}
& \left.\left(\frac{\partial\Psi}{\partial t}
     +\frac{\partial\Psi}{\partial r} \right)\right|_{r = b}
     =
\\
& \frac{1}{b^2}\sum_{j=1}^\ell k_{\ell,j}\int_0^t
      \exp\big(b^{-1}k_{\ell,j}(t-t')\big)\Psi(t',b)dt'.
\end{split}
\end{align}
Here $\{k_{\ell,j}: j = 1,\dots,\ell\}$ are the roots of the
modified cylindrical Bessel function $K_{\ell + 1/2}(x)$,
also known as MacDonald's function. All $k_{\ell,j}$ lie in the
left-half plane. Moreover, the scaled roots $k_{\ell,j}/(\ell+1/2)$
accumulate on a fixed transcendental curve as $\ell$ grows
\cite{AGH,LAU}, so the exponentials 
$\exp\big(b^{-1}k_{\ell,j}t\big)$ tend to decay more quickly 
in time $t >0$ for larger $\ell$.

For the flatspace setting at hand, the Jost solution satisfies
\begin{equation}
v''  - \frac{\ell(\ell+1)}{r^2}v = 0,
\end{equation}
and two appropriate linearly independent solutions are the
following:
\begin{equation}
v_L(r) = r^{\ell+1},\qquad v_R(r) = r^{-\ell}.
\end{equation}
We therefore examine to what extent $v_R(r)$ satisfies 
(\ref{eq:nbc}). Straightforward calculation yields
\begin{align}\label{eq:nbcvr}
\begin{split}
& \left.\left(
\frac{\partial v_R}{\partial r} \right)\right|_{r = b}
     =  - b^{-1} v_R(b) \sum_{j=1}^\ell
       \exp\big(b^{-1}k_{\ell,j}t\big)
\\
& + \frac{1}{b^2}\sum_{j=1}^\ell k_{\ell,j}\int_0^t
      \exp\big(b^{-1}k_{\ell,j}(t-t')\big)v_R(b)dt'.
\end{split}
\end{align}
The function $v_R(r)$ does not satisfy the non-reflecting
condition (\ref{eq:nbc}); however, the violation of
(\ref{eq:nbc}) decays exponentially fast. For 
blackhole perturbations we likewise expect that
$v^\mathrm{axial/polar}_R(\rho)$ violates our radiation
boundary conditions only by exponentially decaying 
terms, and have seen some evidence of this behavior in 
our numerical simulations. 

We have also observed persistent junk solutions when adopting
the Sommerfeld condition at the outer boundary $b$ along with 
impulsive start-up. We differentiate between two scenarios: the 
first involving a detector which is not in causal contact with the 
outer boundary $b$ during the simulation, and a second with the 
detector located at $b$. For the first scenario, the static 
junk solution which develops and persists around the detector 
is precisely $\Psi_\mathrm{Jost}$. For the second, we also
observe a persistent junk solution, but one which is distorted 
from $\Psi_\mathrm{Jost}$ in a boundary layer near $b$. 
Such distortion presumably arises since $\Psi_\mathrm{Jost}$ 
satisfies the outer Sommerfeld condition only up to an 
$O(r^{-\ell-1})$ error term.
%%%%%%%%%%%%%%%%%%%%%%%%%%%%%%%%%%%%%%%%%%%%%%%%%%%%%%%%%%%%%%%
%
%  SECTION 3: CONSEQUENCES OF IMPULSIVE STARTING CONDITIONS
%
%%%%%%%%%%%%%%%%%%%%%%%%%%%%%%%%%%%%%%%%%%%%%%%%%%%%%%%%%%%%%%%
\section{Consequences of impulsive 
         starting conditions}\label{sec:consequences}
\subsection{Inconsistent modeling of the axial 
sector}\label{sec:axialself}
Axial perturbations are described by either the 
Cunningham-Price-Moncrief master function $\Psi^{\mathrm{CPM}}$ 
or the Regge-Wheeler master function $\Psi^{\mathrm{RW}}$. Both 
solve the generic wave equation (\ref{eq:genericwaveeq}) with 
potential (\ref{eq:ReggeWheelerV}).
However, the wave equations for $\Psi^{\mathrm{CPM}}$ and
$\Psi^{\mathrm{RW}}$ have different distributional source 
terms \cite{Martel_CovariantPert,dG_EMRB,Martel_GravWave}. 
As shown in \cite{Martel_CovariantPert}, these master
functions obey
\begin{align}
\label{eqn:CPM-RW-selfconsistency}
\Psi^{\mathrm{RW}} + \half\Pi^\mathrm{CPM} = 0,
\qquad r \neq r_p(t),
\end{align}
and we refer to this formula as the {\em axial consistency 
condition}. For circular orbits this condition becomes
$\Psi^{\mathrm{RW}} - \half\partial_t\Psi^\mathrm{CPM} = 0$,
$r \neq r_p$. We now numerically examine the extent to which 
the axial consistency condition is violated when the master 
functions $\Psi^\mathrm{RW,CPM}$ are obtained with and without 
smoothing.
%*%*%*%*%*%*%*%*%*%*%*%*%*%*%*%*%*%*%*%*%*%*%*%*%*%*%*%*%*%*%*% 
\begin{figure*}
\centering
\includegraphics[height=4.75in]{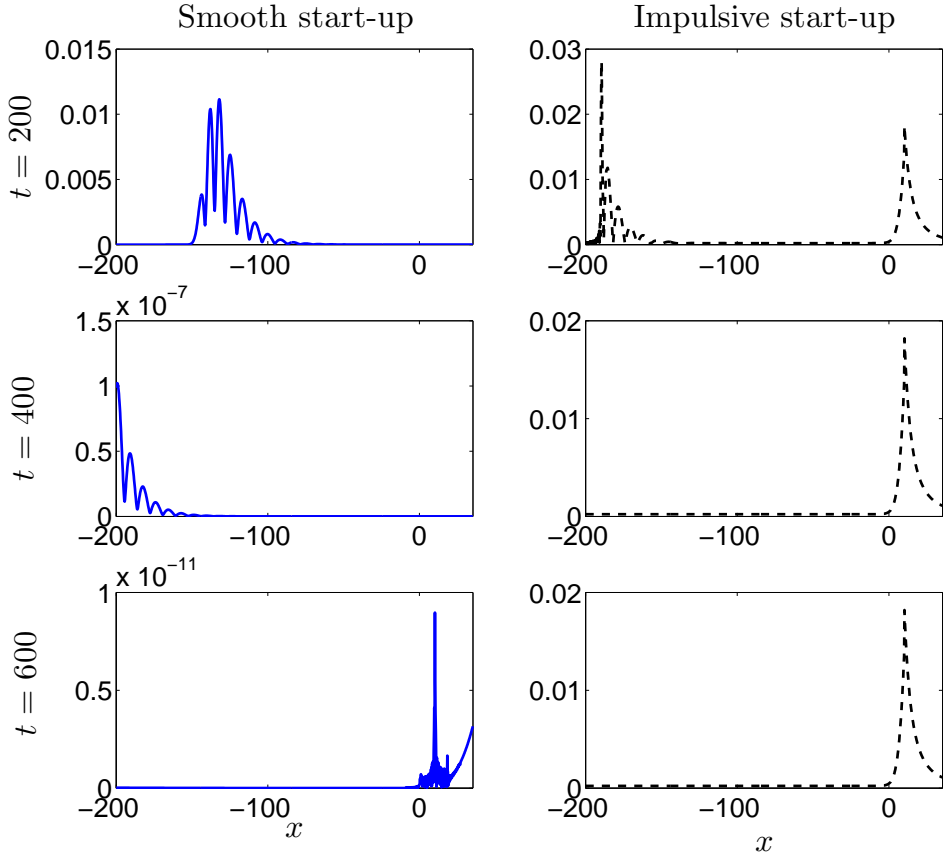}
\caption{{\sc Snapshots of
              $|\Psi^\mathrm{RW} +
              \frac{1}{2}\Pi^\mathrm{CPM}|$
              with and without smoothing.}
The left three panels correspond to smooth start-up
and the right three to impulsive start-up. The times
at the far left correspond to both sets of panels. 
$\Psi^{\mathrm{RW}}$ is of order $10^{-2}$ near $r_p$.}
\label{fig:Circular_RW_CPM_metericPert_smoother}
\end{figure*}
%*%*%*%*%*%*%*%*%*%*%*%*%*%*%*%*%*%*%*%*%*%*%*%*%*%*%*%*%*%*%*% 
%*%*%*%*%*%*%*%*%*%*%*%*%*%*%*%*%*%*%*%*%*%*%*%*%*%*%*%*%*%*%*% 
\begin{figure*}
\centering
\includegraphics[height=3.5in]{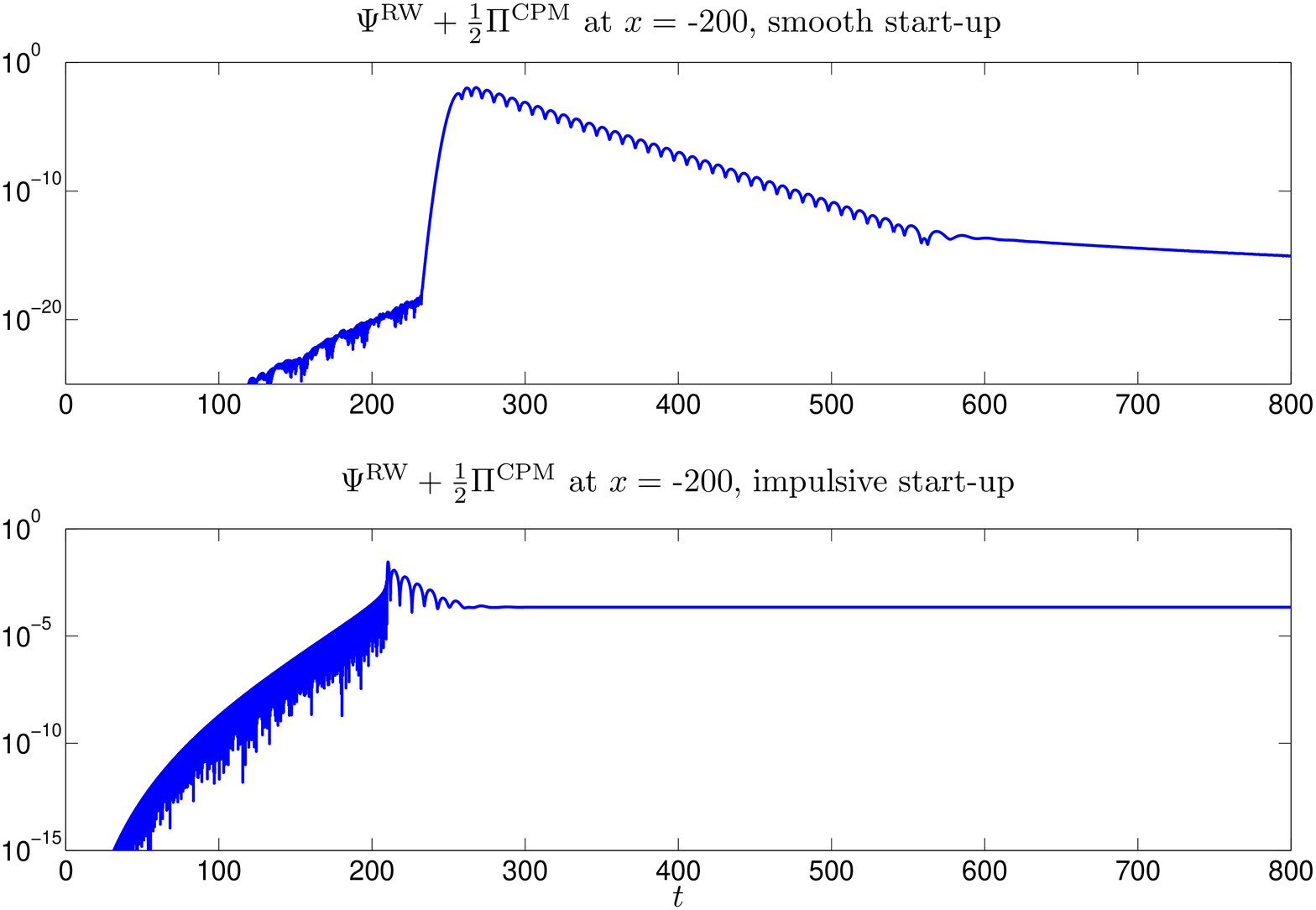}
\caption{{\sc Time series at $x = -200$ for
              $|\Psi^\mathrm{RW} +
              \frac{1}{2}\Pi^\mathrm{CPM}|$
              with and without smoothing.}
$\Psi^{\mathrm{RW}}$ is of order $10^{-4}$ at $x=-200$.}
\label{fig:L2error_vs_time_cirSmooth}
\end{figure*}
%*%*%*%*%*%*%*%*%*%*%*%*%*%*%*%*%*%*%*%*%*%*%*%*%*%*%*%*%*%*%*% 

For all experiments we again enforce Sommerfeld boundary conditions 
at the left physical boundary, and radiation outer boundary conditions 
on the right boundary. Now our smoothing parameters 
are $t_0 = 0$, $\tau = 100$, and $\delta = 0.05$. We compute the 
$(\ell,m)=(2,1)$ metric perturbations for a particle in circular 
orbit initially at $(r,\phi)=(7.9456,0)$. Other parameters 
(cf.~Table \ref{table:parameters}) are $a = -200$, $b = 30 + 2\log(14) 
\simeq 35.28$, $S_L = 22$, $S_R = 3$, $N = 31$, $\Delta t = 0.01$, 
and $t_F = 800$. We first plot 
$|\Psi^\mathrm{RW}+\half\Pi^\mathrm{CPM}|$ at various times. The left 
panels in Fig.~\ref{fig:Circular_RW_CPM_metericPert_smoother} show 
results with smoothing. Although the consistency condition is initially 
violated, the expression eventually relaxes to a small value once the 
dynamical junk has propagated off the domain. The right 
panels in Fig.~\ref{fig:Circular_RW_CPM_metericPert_smoother} show 
result without smoothing. Even at late times violation in the axial
consistency condition is now evident. The plots in 
Fig.~\ref{fig:L2error_vs_time_cirSmooth} depict 
$|\Psi^\mathrm{RW}+\half\Pi^\mathrm{CPM}|$ recorded as a 
time series at $x = -200$. The plot for smooth 
start-up indicates that quasinormal ringing and Price decay tails 
characterize the late-stage dynamical junk, although this 
ringing is suppressed with more smoothing 
(e.~g.~with $\tau = 150$, $\delta = 0.0058$). The plot for 
impulsive start-up suggests that a static Jost junk solution
$\Psi^\mathrm{RW}_\mathrm{impulsive}-\Psi^\mathrm{RW}_\mathrm{smooth}$
persists indefinitely ($\Pi^\mathrm{CPM}$ should be unaffected
by a similar Jost solution in $\Psi^\mathrm{CPM}$). 

\subsection{Contamination of waveforms}

For a given $(\ell,m)$ multipole either read off 
at a finite radius or measured at null infinity 
through an approximate extraction, we can apply standard formulas to
estimate the energy and angular momentum carried away by the 
gravitational waves. We continue to work with the axial perturbations, 
with formulas featuring only CPM and RW masterfunctions. The 
luminosity expressions are the following: 
\cite{Martel_CovariantPert,SopuertaLaguna,Martel_GravWave}
\begin{subequations} \label{eqn:ELflux}
\begin{align}
\begin{split}
\dot{E}_{\ell m}^{\mathrm{CPM}} & = 
\frac{1}{64\pi}
\frac{(\ell+2)!}{(\ell-2)!}
\big|\dot{\Psi}^{\mathrm{CPM}}_{\ell m}\big|^2
\\
\dot{L}_{\ell m}^{\mathrm{CPM}} & = 
\frac{\mathrm{i} m}{64\pi}
\frac{(\ell+2)!}{(\ell-2)!}
\bar{\Psi}_{\ell m}^{\mathrm{CPM}}
\dot{\Psi}^{\mathrm{CPM}}_{\ell m}
\end{split}
\\
\begin{split}
\dot{E}_{\ell m}^{\mathrm{RW}} & =
\frac{1}{16\pi}
\frac{(\ell+2)!}{(\ell-2)!}
\big|\Psi^{\mathrm{RW}}_{\ell m}\big|^2
\\ 
\dot{L}_{\ell m}^{\mathrm{RW}} & = 
\frac{\mathrm{i} m}{16\pi}
\frac{(\ell+2)!}{(\ell-2)!}
\Psi^{\mathrm{RW}}_{\ell m}\int
\bar{\Psi}^{\mathrm{RW}}_{\ell m} dt.
\end{split}
\end{align}
\end{subequations}
In the distant wave-zone we expect 
$\dot{E}_{\ell m}^{\mathrm{CPM}} =
\dot{E}_{\ell m}^{\mathrm{RW}}$
and 
$\dot{L}_{\ell m}^{\mathrm{CPM}} = 
\dot{L}_{\ell m}^{\mathrm{RW}}$. 
However, Sec.~\ref{sec:axialself} has shown that impulsive start-up 
can result in violation of the axial consistency condition 
(\ref{eqn:CPM-RW-selfconsistency}), and such violation in turn 
results in discrepancies between the above luminosity formulas. As 
seen in Sec.~\ref{sec:StaticMasterEqn}, whether simulations are
based on $\Psi^{\mathrm{CPM}}$ or $\Psi^{\mathrm{RW}}$, an impulsive 
start-up generates a Jost junk solution, even at long distances from 
the source. Although dynamical junk is also present, its effect is 
negligible in the wave-zone at late times.

Table~\ref{table:circ_waveformQunt} collects summed luminosities 
for $(\ell,m)=(2,\pm1)$ waveforms. The top set of numbers are
unaveraged and recorded at time $t_F = 2750$, while the bottom set
have been averaged between $t = 2500$ and $t_F = 2500 + 4T_\phi$,
where $T_\phi = 2\pi p^{3/2} \simeq 140.7246$.
Other parameters (cf.~Table \ref{table:parameters}) are 
$a \simeq -190.34$, $b = 1000 + 2\log(499) \simeq 1012.43$, 
$S_L = 30$, $S_R = 150$, $N = 26$, and $\Delta t = 0.038$.
For smoothing we use $\tau = 150$ and $\delta = 0.0058$. 
For circular orbits we expect 
$\langle\dot{Q}_{\mathrm{smooth}}\rangle 
= \dot{Q}_{\mathrm{smooth}}$, 
where brackets denote time averaging for a generic luminosity 
$\dot{Q}$. Relative errors are computed by 
\begin{align}
\dot{Q}_{\mathrm{error}} = 
\frac{\big|\dot{Q}_\mathrm{smooth}   
     -\dot{Q}_\mathrm{impulsive}\big|
    }{\big|\dot{Q}_\mathrm{smooth}\big|}.
\end{align}
For the CPM luminosities computed with smoothing, time 
averaging has little effect. However, it does enhance the accuracy 
of the RW luminosities computed with smoothing. Indeed, 
inspection of the bottom section of 
Table~\ref{table:circ_waveformQunt} shows that the CPM and RW 
entries in the $\dot{Q}_\mathrm{smooth}$ column are in excellent 
agreement.

Relative to the true luminosity which would be recorded at null 
infinity, even the exact $\dot{E}^{\mathrm{CPM}}$ read off at 
$r = 1000$ would have an $O(r^{-1})$ error, but here we have 
viewed the read-off value as the true one. Because 
$\dot{E}^{\mathrm{CPM}}$ is unaffected by the Jost junk solution, 
$\dot{E}^{\mathrm{CPM}}_{\mathrm{error}}$ estimates error 
stemming from both the method (here the Chebyshev scheme) and 
any residual dynamical junk. The other luminosities are affected 
by the Jost junk solution; however, as shown in the Appendix, 
errors which stem from the Jost solution decay faster than $1/r$.
Therefore, these errors should be smaller than the $O(r^{-1})$ 
errors associated with using the read-off luminosities as 
approximations to the ones at null infinity.
\begin{table*}\scriptsize
\begin{tabular}{|l|l|l|l|l|}
\hline
$\dot{Q}$                                     &
$\dot{Q}_\mathrm{smooth}$                     &
$\dot{Q}_\mathrm{impulsive}$                  &
$\dot{Q}_\mathrm{error}$                      \\
\hline
\hline
$\dot{E}^\mathrm{CPM}$                        &
$8.17530620 \times 10^{-7}$                   &
$8.17530623 \times 10^{-7}$                   &
$3.4668 \times 10^{-9}$                       \\
$\dot{E}^\mathrm{RW}$                         &
$8.17530652 \times 10^{-7}$                   &
$8.18248752 \times 10^{-7}$                   &
$8.7838 \times 10^{-4}$                       \\
$\dot{L}^\mathrm{CPM}$                        &
$1.83102415 \times 10^{-5} + \mathrm{i}3.24326408 \times 10^{-14}$ &
$1.82972897 \times 10^{-5} - \mathrm{i}1.28610911 \times 10^{-8}$  &
$9.9685 \times 10^{-4}$                       \\
$\dot{L}^\mathrm{RW}$                         &
$1.83047467 \times 10^{-5} - \mathrm{i}2.16502183 \times 10^{-8}$  &
$1.66825388 \times 10^{-5} + \mathrm{i}8.14152318 \times 10^{-7}$  &
$9.9693 \times 10^{-2}$                       \\
\hline
\hline
$\langle\dot{E}^\mathrm{CPM}\rangle$          &
$8.17530620 \times 10^{-7}$                   &
$8.17530620 \times 10^{-7}$                   &
$2.8376 \times 10^{-10}$                      \\
$\langle\dot{E}^\mathrm{RW}\rangle$           &
$8.17530617 \times 10^{-7}$                   &
$8.17531431 \times 10^{-7}$                   &
$9.9661 \times 10^{-7}$                       \\
$\langle\dot{L}^\mathrm{CPM}\rangle$          &
$1.83102416 \times 10^{-5} - \mathrm{i}1.40467882 \times 10^{-15}$ &
$1.83102416 \times 10^{-5} + \mathrm{i}3.49294212 \times 10^{-14}$ &
$2.0738 \times 10^{-9}$                       \\
$\langle\dot{L}^\mathrm{RW}\rangle$           &
$1.83102415 \times 10^{-5} + \mathrm{i}4.13269715 \times 10^{-13}$ &
$1.82927679 \times 10^{-5} + \mathrm{i}7.05636411 \times 10^{-9} $ &
$1.0292 \times 10^{-3}$                       \\
\hline
\multicolumn{4}{c}{}                            
\end{tabular}
\caption{{\sc $\ell = 2$ luminosities recorded 
at $r = 1000$.} Entries result from addition of 
$m=1$ and $m = -1$ luminosities, and they correspond to 
a circular orbit with $(r,\phi)=(7.9456,0)$ initially.
$\dot{Q}_\mathrm{error}$ as been computed with more 
precision than reported for the table entries.
}
\label{table:circ_waveformQunt}
\end{table*}

\subsection{Self-force measurements}

Over long times the influence of the metric perturbations 
on the particle orbit should significantly affect the gravitational 
waveform \cite{Burko2003}, and realistic waveform computations will
therefore need to include this influence. Incorporation of self-force 
effects constitutes one approach towards modeling this influence.
Because the metric perturbations are discontinuous at the particle, 
self-force calculations typically require a regularization technique.
In the Regge-Wheeler gauge no regularization procedure exists for 
generic orbits; however, direct field-regularization 
\cite{VegaFieldReg,Vega3p1} seems promising. For the restricted 
case of circular orbits, Detweiler has shown how to directly 
calculate certain gauge invariant quantities in the RW gauge 
without regularization \cite{Detweiler}. Detweiler's approach 
obtains the energy luminosity $\dot{E}_{GW}$ and angular momentum 
luminosity $\dot{L}_{GW}$ associated with waves escaping to null 
infinity and down the black hole through local self-force 
calculations performed at the particle,
\begin{equation}\label{eq:EL_evolve}
\dot{E}_p = -\frac{1}{2u^t}u^{\alpha}u^{\beta}
\frac{\partial h_{\alpha \beta}}{\partial t},\qquad
\dot{L}_p = \frac{1}{2u^t}u^{\alpha}u^{\beta}
\frac{\partial h_{\alpha \beta}}{\partial \phi},
\end{equation}
where the perturbation 
$h_{\alpha \beta}$ is reconstructed from $\Psi$ and its 
derivatives \cite{LoustoMetricRecon}. 
Equations (\ref{eq:EL_evolve}) hold for each $(\ell,m)$ 
mode of the metric perturbation.
For perturbations described by the CPM masterfuntion and
with the Regge-Wheeler gauge, the non-zero contributions 
(for each mode) involve
\begin{align}\label{eq:htp_tANDhtp_p}
\begin{split}
\frac{\partial h_{t \phi}}{\partial t}    
       & =  \frac{f}{2}\left(
         r\frac{\partial^2\Psi}{\partial t\partial r}
       +  \frac{\partial\Psi}{\partial t}
                      \right)X_{\phi}
\\
\frac{\partial h_{t \phi}}{\partial \phi} 
       & =  \frac{f}{2}\left(
         r\frac{\partial\Psi}{\partial r}
       +  \Psi
                     \right)X_{\phi\phi}
\end{split}
\end{align}
in a source free region. Here $X_{\phi}$ and $X_{\phi\phi}$ are 
axial vector and tensor spherical harmonics 
\cite{Martel_CovariantPert}. When numerically forming these
expressions, we replace $\partial_t\Psi$ and $\partial_r \Psi$
by $-\Pi$ and $f^{-1}\Phi$. Only when evaluated at the 
particle location will $\dot{E}_p$ and $\dot{L}_p$ be 
related to $\dot{E}_{GW}$ and $\dot{L}_{GW}$.
%*%*%*%*%*%*%*%*%*%*%*%*%*%*%*%*%*%*%*%*%*%*%*%*%*%*%*%*%*%*%*% 
\begin{figure*}
\includegraphics[height=3.0in]{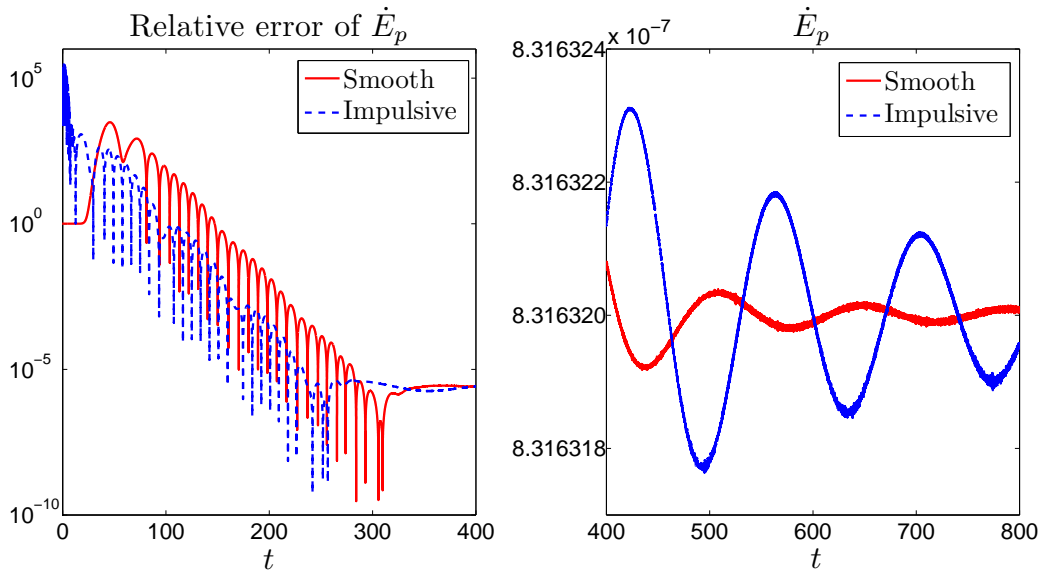}
\caption{{\sc $\dot{E}_p$ time series for summation of $\ell = 2$ and 
$m = \pm1$ modes.} In the right panel the curve corresponding 
to impulsive start-up has the larger amplitude (due 
to small fluctuations this curve does not appear dashed as 
indicated in the legend). 
}
\label{fig:SelfForceTestE}
\end{figure*}
%*%*%*%*%*%*%*%*%*%*%*%*%*%*%*%*%*%*%*%*%*%*%*%*%*%*%*%*%*%*%*% 

We now fix $\tau = 100$ and $\delta = 0.014$ to achieve a 
smooth start-up, run to the final time $t_F = 800$, and pick 
$\Delta t = 0.005$. Other parameters are the same as those in 
Sec.~\ref{sec:axialself}.
%
% Those being  
%
%$a = -200$, $b = 30 + 2\log(14) \simeq 35.2$, 
%$S_L = 22$, $S_R = 3$, $N = 31$.
%
We compute $\dot{E}_p$ and $\dot{L}_p$ 
for $(\ell,m)=(2,\pm1)$ perturbations. Because $\dot{E}_p$ is computed 
with time derivatives of $\Psi^{\mathrm{CPM}}$, the static Jost junk 
solution does not impact its measurement. We therefore expect that
\begin{equation}
\dot{E}_p\big(\Psi^{\ell m}_\mathrm{impulsive} \big) \simeq
\dot{E}_p\big(\Psi_\mathrm{smooth}^{\ell m} \big).
\end{equation}
However, an impulsive start-up 
appears to generate more dynamical junk at late times. 
Figure \ref{fig:SelfForceTestE} depicts $\dot{E}_p$, recorded as a time 
series, for both impulsive and smooth start-ups.  A separate experiment 
based on waveform read-off near the blackhole and waveform extraction at 
the outer boundary determines that the energy carried away by the 
gravitational waves is $\dot{E}_{GW} \simeq 8.3163\times10^{-7}$.
The relative errors in the left panel of 
Fig.~\ref{fig:SelfForceTestE} are computed as 
$|\dot{E}_p-\dot{E}_{\mathrm{GW}}|/\dot{E}_{GW}$, and are 
limited by the accuracy of $\dot{E}_{GW}$. We therefore do not
expect agreement beyond a relative error of $10^{-5}$, although
clearly such error will settle to a constant value.
The time series for both the impulsive and smooth start-up exhibit 
large oscillations which persist until about $t=400$. However, 
beyond $t=400$ the impulsive start-up series shows larger
oscillations. 

\begin{figure*}
\includegraphics[height=2.75in]{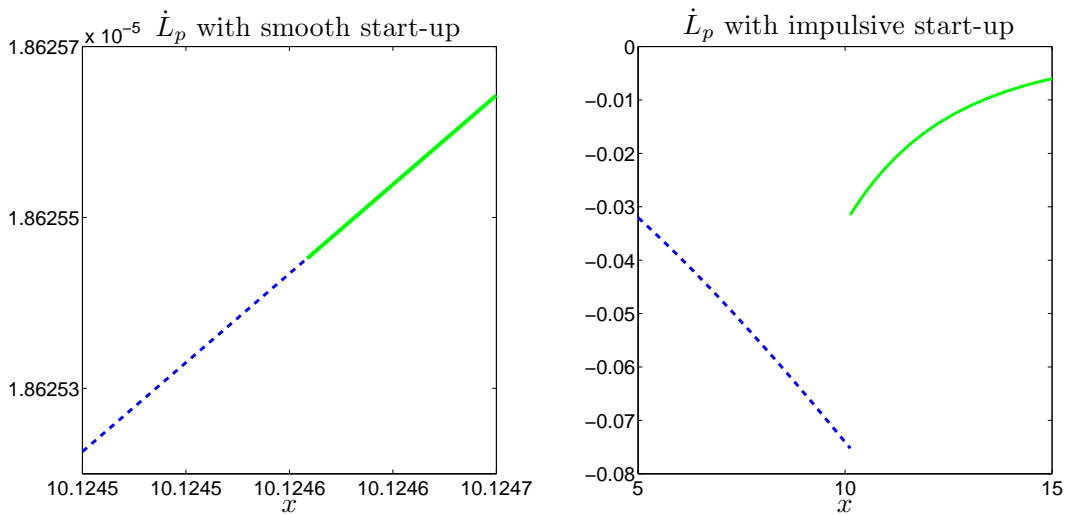}
\caption{{\sc $t=800$ snapshot of real part of $\dot{L}_p$
for $\ell = 2$ and $m = 1$.} The particle is located at the
interface between the two subdomains.
}
\label{fig:SelfForceTestL_snap}
\end{figure*}

$\dot{L}_p$ depends on both $\Psi^{\mathrm{CPM}}$ and its spatial 
derivative $\Phi^{\mathrm{CPM}}$, whence the Jost junk solution will 
impact its self force measurement. With smoothing, the time series 
plot for $\dot{L}_p$ looks similar to one for $\dot{E}_p$ in 
Fig.~\ref{fig:SelfForceTestE}, and is not shown. We note that our 
self-force $\dot{L}_p$ measurement agrees with a separate experiment 
which finds that the angular momentum carried away by gravitational 
waves is $\dot{L}_{GW} \simeq 1.8626\times10^{-5}$. Figure 
\ref{fig:SelfForceTestL_snap} shows that $\dot{L}_p$ is typically 
discontinuous at the particle for an impulsive start-up. Even with 
an impulsive start-up, the $\dot{L}_p$ measurement yields the 
correct value when averaged over an orbital period $T_\phi$, and 
it is continuous across the particle (with the correct value) when 
the particle returns to its initial orbital angle. 
%*%*%*%*%*%*%*%*%*%*%*%*%*%*%*%*%*%*%*%*%*%*%*%*%*%*%*%*%*%*%*%
%*%*%*%*%*%*%*%*%*%*%*%*%*%*%*%*%*%*%*%*%*%*%*%*%*%*%*%*%*%*%*%

These phenomena are a consequence 
of the axial Jost junk solution (\ref{eqn:zeroPsi_general}). 
For $t$ fixed, Eq.~(\ref{eq:EL_evolve}) shows that 
$\dot{L}_p(\Psi)$ depends linearly on $\Psi$. Therefore, 
      $\dot{L}_p
\big(\Psi_{\mathrm{Jost}}^{\ell m} + 
      \Psi_{\mathrm{smooth}}^{\ell m}\big) = 
      \dot{L}_p\big(
      \Psi_{\mathrm{Jost}}^{\ell m} \big) + 
      \dot{L}_p\big(
      \Psi_{\mathrm{smooth}}^{\ell m}\big)$,
so we can focus on 
     $\dot{L}_p\big(\Psi_{\mathrm{Jost}}^{\ell m} \big)$ alone.
The expressions (\ref{eqn:constantsEqn2}) 
for $C_{L,R}$ are linear in $F(0,r_p)$, which is in turn proportional
to the conjugate of an axial vector spherical harmonic $X_\phi$ 
\cite{Martel_CovariantPert}. 
Motivated by this observation, we ``factor off" the conjugate, 
writing $\Psi_{\mathrm{Jost}}^{\ell m}=
\eta_{\ell}(x)\bar{X}^{\ell m}_{\phi}(\phi_0)$,
where $\phi_0$ is the particle's initial orbital angle and 
$\eta_{\ell}(x)$ is a real discontinuous function solely of $x$.
The expression (\ref{eq:EL_evolve}) for 
$\dot{L}_p$ involves $\partial h_{t\phi}/\partial\phi$, which by 
(\ref{eq:htp_tANDhtp_p}) is proportional to $X_{\phi\phi}$. In the 
equatorial plane 
    $X_{\phi\phi}^{\ell m} 
    = \partial_{\phi}X_{\phi}^{\ell m} 
    = \mathrm{i} m X_{\phi}^{\ell m}$, 
and we conclude that
    $\dot{L}_p\big(\Psi_{\mathrm{Jost}}^{\ell m} \big) 
    =\mathrm{i} m \xi_{\ell}(x) 
    \bar{X}^{\ell m}_{\phi}(\phi_0) 
    X_{\phi}^{\ell m}(\phi_p(t))$, 
where $\xi_{\ell}(x)$ is a real discontinuous function solely of 
$x$. Therefore, when the particle returns to its initial position 
(that is, when $\phi_p(t) = \phi_0$), the value of 
$\dot{L}_p\big(\Psi_{\mathrm{Jost}}^{\ell m} \big)$
is pure imaginary and
    $\dot{L}_p\big(\Psi_{\mathrm{Jost}}^{\ell m} \big) 
    +\dot{L}_p\big(\Psi_{\mathrm{Jost}}^{\ell, -m} \big) = 0$.
For perturbations generated by a particle in circular orbit, we 
have seen that 
$\Psi_\mathrm{impulsive}^{\ell m}
\simeq 
\Psi_\mathrm{Jost}^{\ell m} +
\Psi_\mathrm{smooth}^{\ell m}$ to high accuracy.
Combination of this expression and the above arguments for axial
perturbations then gives
\begin{equation}
\sum_{|m| \leq \ell} \dot{L}_p\big(\Psi_\mathrm{impulsive}^{\ell m}\big) 
\simeq
\sum_{|m| \leq \ell} \dot{L}_p\big(\Psi_\mathrm{smooth}^{\ell m} \big),
\end{equation}
when $\phi_p(t) = \phi_0$.
Moreover, one finds
$\big\langle
 \dot{L}_p\big(\Psi_{\mathrm{Jost}}^{\ell m} 
 \big)
 \big\rangle = 0$ for time averaging over an orbital period $T_\phi$.
%*%*%*%*%*%*%*%*%*%*%*%*%*%*%*%*%*%*%*%*%*%*%*%*%*%*%*%*%*%*%*%
\begin{figure*}
\centering
\includegraphics[height=4.5in]{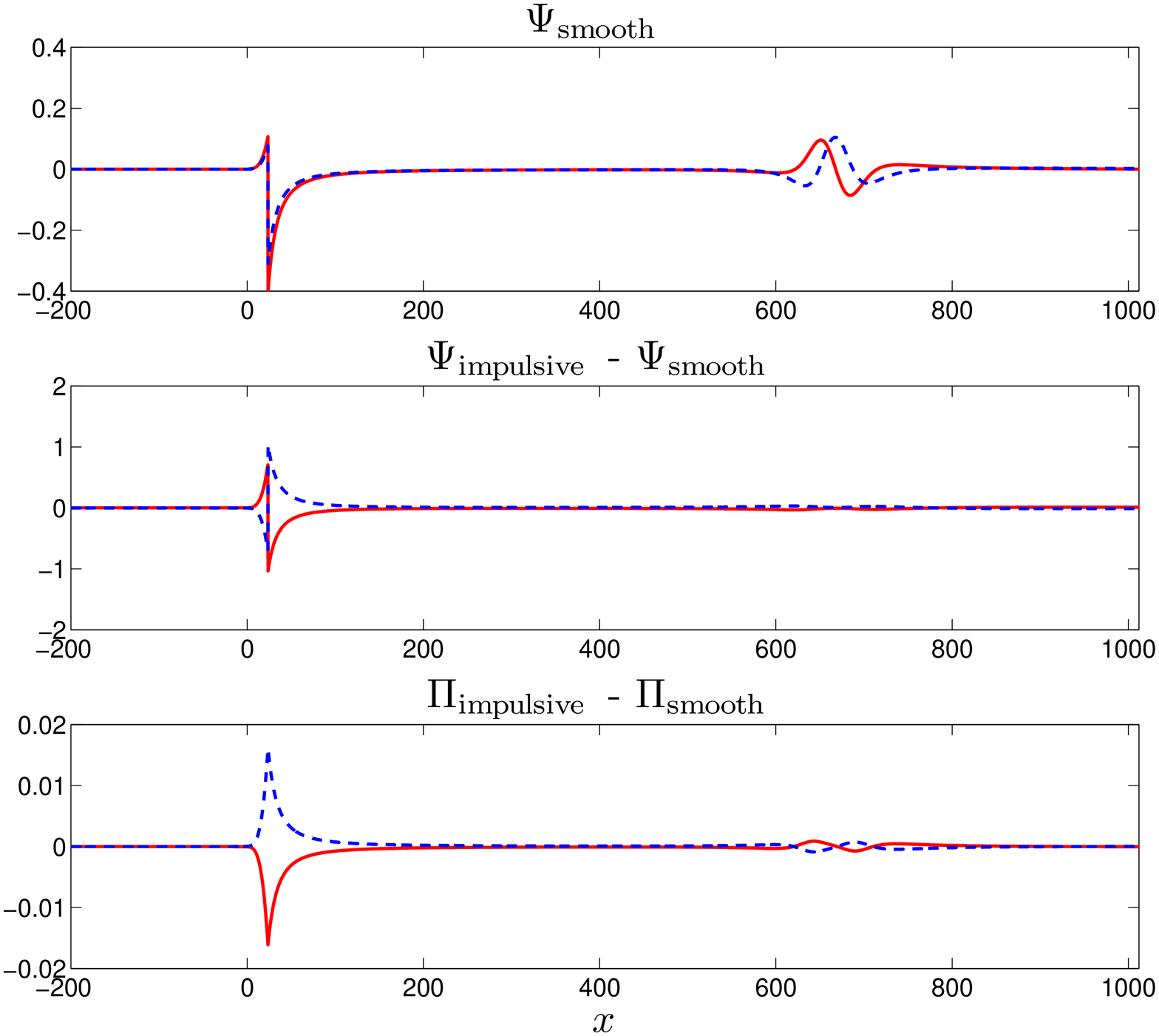}
\caption{{\sc Difference between CPM fields with and without
              smoothing for an eccentric orbit.} Here we
              plot both real (dashed) and imaginary (solid)
              parts at $t_F = 3000$.
        }
\label{fig:Ecc_RW_CPM_metricPert}
\end{figure*}
%*%*%*%*%*%*%*%*%*%*%*%*%*%*%*%*%*%*%*%*%*%*%*%*%*%*%*%*%*%*%*%

\subsection{Consequences for eccentric orbits: preliminary results}
This section considers a particle in the eccentric orbit 
described in Section IV.B.2 of \cite{dG_EMRB}. In the notations 
of that reference the orbit's eccentricity and semi-latus rectum 
are $(e=0.76412402,p=8.75456059)$, and we choose $\chi=0.2$ and 
$\phi=\pi/4$ to fix the particle's initial position. We simulate the 
resulting $(\ell,m)=(2,1)$ perturbation with (cf.~Table 
\ref{table:parameters}) $a = -200$, $b = 1012.43$, $S_L = 22$,
$S_R = 100$, $N = 31$, $\Delta t = 0.02$, and $t_F = 3000$. We 
again take $\tau = 150$, $\delta = 0.0058$ as the smoothing 
parameters. Since $e \neq 0$, we 
use a coordinate transformation to keep the particle at a fixed 
location between subdomains (see \cite{dG_EMRB} for details). 
Before making comparisons, we first interpolate all fields onto 
a uniform $x$--grid (tortoise coordinate) with 6063 points.

Fig.~\ref{fig:Ecc_RW_CPM_metricPert} shows
the difference between fields for smooth and impulsive start-ups.
The two numerical solutions are clearly different, although for
the case of eccentric orbits we have no analytical understanding
of the resulting ``junk solution"\footnote{At present, we 
are uncertain if the generated junk solution fulfills the formal 
definition of a Jost solution. Thus, in the context of eccentric 
orbits we simply refer to the persistent solution as the 
``junk solution".} presumably seeded by impulsive start-up. 
Empirically, we find that this solution satisfies
\begin{subequations}
\begin{align}
 &\ljump \Psi_{\mathrm{junk}}\rjump(t) = -
 \ljump\Psi_{\mathrm{analytic}}\rjump(0) \\ 
 &\ljump \Phi_{\mathrm{junk}} \rjump(t) = 0 \\
 &\ljump \Pi_{\mathrm{junk}} \rjump(t) = 0,
\end{align}
\end{subequations}
where $\ljump\Psi_{\mathrm{analytic}}\rjump(t) =
f_p(t)F(t,r_p(t))/(f_p^2(t)-\dot{r}^2_p(t))$ in terms of
$f_p(t) = f(r_p(t))$. See \cite{dG_EMRB} for a derivation of 
the analytical jump. These time independent jump conditions 
are the same as for the circular orbit $\Psi_{\mathrm{Jost}}$ 
solution.  
With our choice of numerical parameters the axial consistency 
condition is satisfied to better than a $1\times 10^{-6}$ 
relative error throughout the entire domain for a smooth start-up. 
For an impulsive start-up the condition is violated to the 
order of the solution itself. We conclude that, as for circular
orbits, the junk solution generated by an impulsive start-up 
leads to an inconsistent modeling of the axial sector.

Table \ref{table:ecc_waveformQunt} collects energy and angular
momentum luminosities. These luminosities have been averaged
from $t = 1700$ to $t_F = 1700 + 4T_r$, where
$T_r \simeq 780.6256$ is the radial period (see \cite{dG_EMRB}
for further details). Unlike the circular orbit case, the
discrepancy between waveforms corresponding to smoothly and
impulsively started fields may be larger than usual $O(1/r)$
error associated with read-off at a finite radial location
rather than infinity. Moreover, the junk solution would seem
determined by the initial orbital parameters. Indeed, the
values $\dot{Q}_\mathrm{impulsive}$ and errors 
quoted in our table strongly depend upon such choices. 

\begin{table*}\scriptsize
\begin{tabular}{|l|l|l|l|l|}
%\hline
%\multicolumn{4}{|c|}{Quantities Calculated From 
%Waveforms Recorded at $r^*=1000M$}    \\
%\hline
\hline
$\dot{Q}$                            &
$\dot{Q}_{\mathrm{smooth}}$                     &
$\dot{Q}_{\mathrm{impulsive}}$                  &
$\dot{Q}_{\mathrm{error}}$                      \\
\hline
\hline
$\langle\dot{E}^{\mathrm{ZM}}_{2,2}\rangle$ & 
$1.559917 \times 10^{-4}$     &
$1.559484 \times 10^{-4}$        &
$2.775789 \times 10^{-4}$        \\
$\langle\dot{E}^{\mathrm{CPM}}_{2,1}\rangle$ & 
$1.153983 \times 10^{-6}$    &
$1.236758 \times 10^{-6}$    &
$7.172983 \times 10^{-2}$        \\
$\langle\dot{E}^{\mathrm{RW}}_{2,1}\rangle$ & 
$1.153983 \times 10^{-6}$    &
$1.872073 \times 10^{-6}$    &
$6.222709 \times 10^{-1}$        \\
\hline
$\langle\dot{E}^{\mathrm{CPM}}_{2,1}\rangle + 
\langle\dot{E}^{\mathrm{ZM}}_{2,2}\rangle$ & 
$1.571457 \times 10^{-4}$    &
$1.571852 \times 10^{-4}$    &
$2.512000 \times 10^{-4}$        \\
\hline
\hline
$\mathrm{Re}\langle\dot{L}^{\mathrm{ZM}}_{2,2}\rangle$ & 
$2.078556 \times 10^{-3}$    &
$2.076811 \times 10^{-3}$        &
$8.395251 \times 10^{-4}$        \\
$\mathrm{Re}\langle\dot{L}^{\mathrm{CPM}}_{2,1}\rangle$ & 
$1.441737 \times 10^{-5}$    &
$1.537876 \times 10^{-5}$        &
$6.668276 \times 10^{-2}$        \\
$\mathrm{Re}\langle\dot{L}^{\mathrm{RW}}_{2,1}\rangle$ & 
$1.441749 \times 10^{-5}$    &
$1.662726 \times 10^{-5}$        &
$1.532701 \times 10^{-1}$        \\
\hline
$\mathrm{Re}\langle\dot{L}^{\mathrm{CPM}}_{2,1}\rangle 
+ \mathrm{Re}\langle\dot{L}^{\mathrm{ZM}}_{2,2}\rangle$ & 
$2.092973 \times 10^{-3} $    &
$2.092190 \times 10^{-3}$        &
$3.744004 \times 10^{-4}$        \\
\hline
\multicolumn{4}{c}{}           \\
\multicolumn{4}{c}{}           \\

\end{tabular}
\caption{{\sc $\ell = 2$ luminosities for a 
particle with an orbit given by $(e=0.76412402,p=8.75456059)$.}
Entries result from the addition of
$|m|$ and $-|m|$ luminosities.
}
\label{table:ecc_waveformQunt}
\end{table*}

%%%%%%%%%%%%%%%%%%%%%%%%%%%%%%%%%%%%%%%%%%%%%%%%%%%%%%%%%%%%%%%
%
%  SECTION 4: CONCLUSIONS
%
%%%%%%%%%%%%%%%%%%%%%%%%%%%%%%%%%%%%%%%%%%%%%%%%%%%%%%%%%%%%%%%
\section{Conclusions}\label{sec:conclusions}
A number of time-domain methods exist for solving 
Eq.~(\ref{eq:genericwaveeq}) as an initial boundary value 
problem, including those described in \cite{
Barack2009Rev,
Martel_GravWave,
LoustoScheme,
SopuertaLaguna,
JungKhannaNagle,
CanizaresSopuerta,
CanizaresSopuerta1}.
These methods vary in both the underlying numerical
scheme (e.g.~finite difference, finite element, pseudospectral,
and spectral) as well as their treatment of the 
distributional source terms (e.g.~Gaussian representation, 
analytic integration, domain matching). Numerical simulation
of metric perturbations may also involve other choices
(e.g.~gauge, number of spatial dimensions, choice of 
numerical variables). Moreover, similar time-domain methods exist 
for solving the forced Teukolsky equation describing 
particle-driven perturbations of the Kerr geometry (see for 
example Refs.~\cite{Aleman,ScottHughes_Adiabatic,SundararajanGaussin}).
For all of these methods, the issue of impulsive start-up would
seem pertinent, although clearly we cannot examine each method.
Nevertheless, we now attempt to provide at least partial insight 
into the ubiquity of static junk solutions.

%*%*%*%*%*%*%*%*%*%*%*%*%*%*%*%*%*%*%*%*%*%*%*%*%*%*%*%*%*%*%*% 
\begin{figure}
\centering
% trim=l b r t
\includegraphics[clip=true,height=2.8in,trim = 25mm 0 25mm 0]{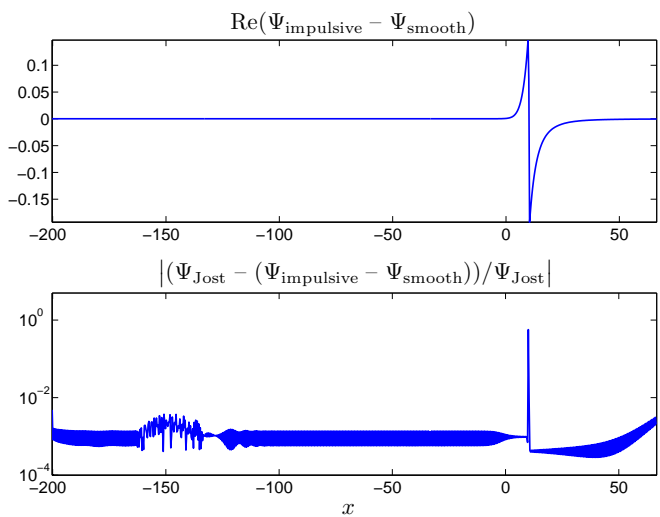}
\caption{{\sc
Difference between smoothly and impulsively started
fields using a finite-difference method.} As in Subsection
\ref{sec:StaticMasterEqn}, we consider $\Psi^\mathrm{CPM}$
for $\ell = 3$ and $m = 2$. The bottom plot depicts the
relative error between the numerical and analytical Jost
solutions. 
}
\label{fig:FiniteDifference}
\end{figure}
%*%*%*%*%*%*%*%*%*%*%*%*%*%*%*%*%*%*%*%*%*%*%*%*%*%*%*%*%*%*%*% 

As mentioned earlier,
the results and observations of this paper have been independently
confirmed with each of our two numerical methods: the nodal 
Legendre dG and Chebyshev schemes. However, as these schemes are
rather similar, we now briefly consider a finite-difference 
scheme for solving (\ref{eq:firstordersys}), based on fourth, 
sixth, and eighth order stencils for the spatial derivatives. 
To stabilize sixth and eighth order stencils, we have followed
Ref.~\cite{HagstromHagstrom}. Furthermore, we replace the Dirac 
delta functions in (\ref{eq:firstordersys}) by narrow Gaussians. 
Precisely, for $\sigma = 0.1$ we make the replacement
\begin{equation}
J(x,t)\delta(x-x_p) 
\rightarrow 
J(x,t)
\frac{1}{\sqrt{2\pi}\sigma}
\mathrm{exp}\left(-\frac{(x-x_p)^2}{2\sigma^2}\right)
\end{equation}
for both the $J_1$ and $J_2$ terms in (\ref{eq:firstordersys}).
Analytic expressions for $J_1$ and $J_2$ are readily computed with 
Eq.~(28) from Ref.~\cite{dG_EMRB}. With essentially 
the same experimental set-up described in Subsection 
\ref{sec:StaticMasterEqn}, we repeat that experiment using 4000
points and sixth order spatial differences. The results, shown
in Fig.~\ref{fig:FiniteDifference}, clearly indicate the 
presence of a static ``Jost junk" solution. A larger choice 
for $\sigma$ gives rise to a rounder transition near the 
particle. 
However, the following shows that such 
contamination is not a generic feature. 
For circular orbits, our system (\ref{eq:firstordersys}) becomes
\begin{align}\label{eq:nodeltaprime}
\begin{split}
\partial_t \Psi & = -\Pi \\
\partial_t \Pi  & = -\partial_x \Phi 
                    + V(r) \Psi 
                    + J_1 \delta (x-x_p)\\
\partial_t \Phi & = -\partial_x \Pi  
                    + J_2 \delta (x-x_p), \\
\end{split}
\end{align}
where the time-dependent jump factors are $J_1 = [[\Psi_x]]$ 
and $J_2 = -[[\Psi_t]]$. We introduce a variable 
$\tilde{\Phi}$ obeying
\begin{equation}
\Phi = \tilde{\Phi} - [[\Psi]]\delta (x - x_p),
\end{equation}
so that the system formally becomes
\begin{align}\label{eq:deltaprime}
\begin{split}
\partial_t \Psi & = -\Pi \\
\partial_t \Pi  & = -\partial_x \tilde{\Phi} + V(r) \Psi 
                    + J_1 \delta (x-x_p) 
                    + J_3 \delta'(x - x_p)\\
\partial_t \tilde{\Phi} & = -\partial_x \Pi,\\
\end{split}
\end{align}
where $J_3 = [[\Psi]] = F(t,r_p)/f_p$. If we now replace the
$\delta$,$\delta'$ terms in the new system by appropriate 
Gaussians, {\em then we do not observe a persistent Jost junk 
solution} when trivial initial conditions are supplied
(neither in finite-difference nor dG simulations).

Persistent junk solutions arise from the combination
of inconsistent initial data and the distributional forcing 
terms which define the EMRB model. In particular, 
we observe that development of a Jost junk solution depends on
how the distributional forcing is treated rather than the 
underlying numerical method. Therefore, whether or not
they contaminate simulations should be considered on a 
case-by-case basis. Domain matching approaches which 
enforce jump conditions without approximation (considered in this 
paper) exhibit a Jost junk solution in the absence of smooth start-up.
With first order variables such approaches correspond to system 
(\ref{eq:nodeltaprime}) rather than (\ref{eq:deltaprime}).
Treatment of system (\ref{eq:deltaprime}) with Gaussian representations
for $\delta$,$\delta'$ exhibits no persistent junk solution, 
although such an approach necessarily introduces large method error
relative to the exact distributional model and features variables
with $\delta$-like behavior near the ``particle" (Gaussian peak).
The issue of a static junk solution for schemes which discretize 
the second order equation (\ref{eq:genericwaveeq}) deserves further 
consideration, although, if present, then the particular Jost junk 
solution observed in this paper would likely be of 
relevance.\footnote{For a static solution to have gone unnoticed, 
it would seem reasonable to expect decay as either 
$r \rightarrow 2M^+$ or $r \rightarrow \infty$. Such solutions 
will necessarily be 
discontinuous, and presumably such discontinuities could only 
``hide" at the particle, requirements that fix the form of 
the static solution up to the constants $C_L$ and
$C_R$ introduced in Section \ref{sec:StaticMasterEqn}.}

We have shown that impulsive starting conditions are inadequate 
for time-domain modeling of extreme mass ratio binaries. Such 
conditions result in more dynamical junk, evident in 
self-force calculations, and potentially a static Jost junk solution 
which persists indefinitely. Although each effect is small compared to
the physical solution, such systematic errors will 
corrupt studies which require high accuracy. 
For example, computation of waveforms accurate to second 
order in the mass ratio requires reconstruction of the first order 
perturbations. Since these first order terms act as sources for 
the wave equations describing the second order masterfunctions, 
the presence of a Jost junk solution will affect second order 
waveforms. Circular orbits far from the massive central object
(of potential relevance for the quasi-circular phase of 
inspiral) are similarly impacted by the Jost junk solution. 
Eq.~(\ref{eqn:constantsEqn2}) indicates that the magnitude of a
polar-mode static junk solutions does not decay as $r_p$ becomes 
large (compare with Eqs.~(C5a) and (C6e) from \cite{dG_EMRB}).
However, such decay is present in the axial case 
(compare with (C8a) and (C9c) of \cite{dG_EMRB}).
When studying eccentric orbits, errors arising from 
the persistent junk solution appear to corrupt studies requiring 
even modest accuracy. Minimization of dynamical and Jost junk by 
smoothing the source terms during start-up will improve waveform 
templates and self-force techniques with minimal computational 
and human effort.

%%%%%%%%%%%%%%%%%%%%%%%%%%%%%%%%%%%%%%%%%%%%%%%%%%%%%%%%%%%%%%%
%
%  SECTION 5: ACKNOWLEDGMENTS
%
%%%%%%%%%%%%%%%%%%%%%%%%%%%%%%%%%%%%%%%%%%%%%%%%%%%%%%%%%%%%%%%
\section{Acknowledgments}\label{sec:acknowledgments}
We thank S.~Detweiler for discussions and correspondence,
L.~Barack for comments offered at the 12\underline{th}
Capra Meeting on Radiation Reaction, and C.~Galley
and M.~Tiglio for answering our questions concerning
effective field theory approaches. We gratefully 
acknowledge funding through NSF grant PHY 0855678
to the University of New Mexico and 
DMS 0554377 and DARPA/AFOSR FA9550-05-1-0108 to Brown University.

%%%%%%%%%%%%%%%%%%%%%%%%%%%%%%%%%%%%%%%%%%%%%%%%%%%%%%%%%%%%%%%
%
%  APPENDIX
%
%%%%%%%%%%%%%%%%%%%%%%%%%%%%%%%%%%%%%%%%%%%%%%%%%%%%%%%%%%%%%%%
\appendix
\section*{Appendix: Time-independent master equations}
\subsection{Regge-Wheeler equation}
Subject to the {\em Ansatz} that the solution $v$ is 
time-independent and in terms of the dimensionless 
variable $\rho = (2M)^{-1}r$, the homogeneous Regge-Wheeler 
equation is \cite{Donninger2009}
\begin{equation}
- \left(1-\frac{1}{\rho}\right)v'' 
- \frac{1}{\rho^2} v' + 
  \left[\frac{\ell(\ell+1)}{\rho^2} 
+ \frac{\kappa}{\rho^3}\right]v = 0,
\end{equation}
where $\kappa = 1-\jmath^2$ in terms of the spin 
$\jmath = 0,1,2$. For gravitational perturbations 
$\jmath = 2$, but we leave $\jmath$ unspecified 
for the time being. Expressing the 
equation in the form
\begin{align}
\begin{split}
& v'' + P(\rho)v' + Q(\rho) v = 0
\\
& P(\rho) = \frac{1}{\rho(\rho-1)},
\qquad Q(\rho) = -\frac{\ell(\ell+1)\rho 
                 + \kappa}{\rho^2(\rho-1)},
\end{split}
\end{align}
we find that it has regular singular points at $0$, $1$, 
and $\infty$, as well as the associated Riemann-Papperitz 
symbol \cite{MathewsWalker}
\begin{equation}
v = P\left\{\begin{array}{cccc}
0        & 1 &   \infty     &      \\
1+\jmath & 0 &  -(\ell + 1) & ; \rho \\
1-\jmath & 0 &   \ell       &
\end{array}\right\}.
\end{equation}
To obtain the standard normal form, we let 
$v = \rho^{1+\jmath} u$, so that
\begin{equation}
u = P\left\{\begin{array}{cccc}
0        & 1 &   \infty            & \\
0        & 0 &  -\ell +\jmath      & ; \rho \\
-2\jmath & 0 &   \ell + \jmath + 1 &
\end{array}\right\},
\end{equation}
where $u$ satisfies the hypergeometric equation
\begin{equation}
\rho(1-\rho) u'' + [c - (a+b+1)\rho]u' - ab u = 0,
\end{equation}
with $a = -\ell + \jmath$, $b = \ell + \jmath + 1$, 
and $c =  1+2\jmath$. As one of the two linearly
independent solutions based at $\rho = \infty$ (chosen
to be the second), we may take
\begin{equation}
u_2(\rho) = \rho^{-\ell - \jmath -1}
{}_2 F_1(\ell + \jmath +1,\ell - \jmath +1; 2(\ell + 
1);\rho^{-1}).
\end{equation}
Expressed in terms of the original dependent variable, 
$v_2 = \rho^{1+\jmath}u_2$, this solution is our axial/right 
solution $v_2(\rho) = v^\mathrm{axial}_R(\rho)$ given in 
(\ref{eqn:axialvLvR}b).
%
%  rho = 0: p0 = -1, q0 = \kappa
%  a(a-1) - a + kap = 0 ==> a^2 -2a + kap = 0
%
%
%  rho = 1, p0 = 1, q0 = 0
%  a(a-1) + a + 0 = 0 ==> a^2 = 0.
%
%  rho = inf: p0 = 2, q0 = -ell(ell+1)
%  a^2 + a - ell(ell+1) = 0
%
%  1 + 4ell^2 + 4ell = (2ell+1)^2 
To obtain series solutions based at $1$ which are 
nevertheless valid on $(1,\infty)$, we follow
Leaver \cite{Leaver} and consider the transformation 
$\eta = (\rho-1)/\rho$. Then with $w(\eta) = v(1/(1-\eta))$, 
we get
\begin{align}
\begin{split}
& w'' + \mathcal{P}(\eta) w' + \mathcal{Q}(\eta)w = 0
\\
& \mathcal{P}(\eta) = \frac{1-3\eta}{\eta(1-\eta)},\quad
\mathcal{Q}(\eta) = - 
\frac{\ell(\ell+1)+ \kappa (1-\eta)}{\eta(1-\eta)^2},
\end{split}
\end{align}
which has the $P$-symbol
\begin{equation}
w = P\left\{\begin{array}{cccc}
0        & 1          & \infty    &       \\
0        & -(\ell+1)  & 1+\jmath  &  ; \eta \\
0        &  \ell      & 1-\jmath  &
\end{array}\right\}.
\end{equation}
Now let $w = (\eta-1)^\ell y$ so that
\begin{equation}
y = P\left\{\begin{array}{cccc}
0        & 1             & \infty         &       \\
0        & 0             & 1+\ell+\jmath  &  ; \eta \\
0        & -(2\ell+1)    & 1+\ell-\jmath  &
\end{array}\right\}
\end{equation}
solves
\begin{equation}
\eta(1-\eta) y'' + [C - (A+B+1)\eta]y' - AB y = 0,
\end{equation}
with $A = \ell-\jmath+1$, $B = \ell+\jmath+1$, and
$C = 1$. Therefore, we choose $v_1(\rho) =
v^\mathrm{axial}_L(\rho)$ given in (\ref{eqn:axialvLvR}a)
as both a first linearly independent solution and 
the axial/left one of interest.

\subsection{Zerilli equation}
In dimensionless form, the time--independent Zerilli 
equation is
\begin{align}
\begin{split}
& -\left(1-\frac{1}{\rho}\right)v'' - \frac{1}{\rho^2} v' 
\\
& + \left[\frac{8 n^2 (n+1)\rho^3 + 12 n^2\rho^2 + 18 
n \rho + 9}{\rho^3(2 n \rho + 3)^2}\right]v = 0,
\label{eq:notimeZerilli}
\end{split}
\end{align}
again where $n = \frac{1}{2}(\ell - 1)(\ell + 2)$. In standard
form, the equation is
\begin{align}
\begin{split}
& v'' + P(\rho)v' + Q(\rho)v = 0,\quad 
P(\rho) = \frac{1}{\rho(\rho-1)}
\\
& 
Q(\rho) = -\left[\frac{8 n^2 (n+1)\rho^3 + 12 n^2\rho^2 + 
18 n \rho + 9}{\rho^2(\rho-1)(2 n \rho + 3)^2}\right].
\end{split}
\end{align}
This equation has regular singular points at $0$, $1$, $\infty$,
and $-3/(2 n)$, with the following associated pairs of
indicial exponents: $\{1,1\}$, $\{0,0\}$, $\{\ell,-(\ell+1)\}$, 
$\{2,-1\}$.
%%  rho           p0              q0   	       	exponents
%%       	       	       	       	       	1, 1
%%  0      	 -1	       	  1            
%%
%%  1      	  1	       	  0   	        0, 0
%%
%% -3/(2n)	  0	       	 -2    	       	2, -1
%%
%%  inf    	  2	       	 -2(n+1)       -1/2 \pm sqrt(9 + 8 n)/2
%%
%%  now 9 + 8 n = 9 + 4 (ell-1)(ell+2) 
%%              = 4 ell^2 + 4 ell + 1 
%%              = (2 ell + 1)^2
%%
%%  so at inf the exponents are -1/2 \pm (2ell+1)/2 = ell, -(ell+1)
The general second order homogeneous ODE with regular singular points 
at $z_0$, $z_1$, $z_2$, and $\infty$ has the form 
$y'' + R(z) y' + S(z)y = 0$, with
\begin{align}
\begin{split}
R(z) & = \frac{A_0}{z-z_0}
     + \frac{A_1}{z-z_1}
     + \frac{A_2}{z-z_2}
\\
S(z) & = \frac{B_0}{(z-z_0)^2}
     + \frac{B_1}{(z-z_1)^2}
     + \frac{B_2}{(z-z_2)^2}
\\
     & + \frac{C_0}{z-z_0}
     + \frac{C_1}{z-z_1}
     + \frac{C_2}{z-z_2},
\end{split}
\label{claimedPQ}
\end{align}
where the $A_i$, $B_i$, and $C_i$ are all constants subject
to $C_0 + C_1 + C_2 = 0$ and the requirement that for each
$i = 0,1,2$ at least one member of the triple $A_i$, $B_i$,
and $C_i$ must be nonzero (for otherwise $z_i$ would be a
ordinary point). By expressing all constants $A_i$, $B_i$, 
$C_i$ except $C_0$ in terms of the indicial exponents 
$\big\{\{\lambda_k, \lambda_k'\} : k=0,1,2,\infty\big\}$, 
we find
\begin{align}
\begin{split}
R(z) & =
        \frac{1-\lamzer-\lamzerp}{z - z_0}
     + \frac{1-\lamone-\lamonep}{z - z_1}
     + \frac{1-\lamtwo-\lamtwop}{z - z_2}
\\
S(z) & =
           \frac{\lamzer\lamzerp}{(z - z_0)^2}
          +\frac{\lamone\lamonep}{(z - z_1)^2}
          +\frac{\lamtwo\lamtwop}{(z - z_2)^2}
\\
     & +
            \frac{\laminf\laminfp
           - \lamzer\lamzerp
           - \lamone\lamonep
           - \lamtwo\lamtwop}{(z-z_1)(z-z_2)}
\\
          & + \frac{C_0(z_0-z_1)(z_0-z_2)}{(z-z_0)
                   (z-z_1)(z-z_2)}\, .
\end{split}
\label{formofPQ}
\end{align}
Here $-C_0$ is the {\em accessory parameter} \cite{SlavyanovLay},
and the generalized Riemann 
scheme \cite{SlavyanovLay} associated with the equation is
\begin{equation}
\left[\begin{array}{ccccl}
1         & 1         & 1         & 1             &     \\
z_0	  & z_1       & z_2	  &\infty         & ; z \\
\lamzer  & \lamone  & \lamtwo  &\laminf  & ; -C_0 \\
\lamzerp & \lamonep & \lamtwop &\laminfp &
\end{array}\right].
\end{equation}
The notation is similar to the $P$-symbol, but also indicates
the type of singular points in the first row (regular singular
points are indicated by a 1). We find the scheme
\begin{equation}
\left[\begin{array}{ccccl}
1         & 1         & 1   	         & 1             &        \\
0         & 1         & -3/(2n) 	 &\infty         & ; \rho \\
1         & 0         & 2                & -(\ell+1)     & ; 0    \\
1         & 0         & -1               & \ell          &
\end{array}\right].
\end{equation}
for the specific case of the time--independent Zerilli equation 
(\ref{eq:notimeZerilli}).

Upon transforming the ODE specified by (\ref{formofPQ}) to normal 
form, we find the new accessory parameter
\begin{align}
\begin{split}
q = -C_0
& +\frac{\lamzer(\lamonep-1)+\lamone(\lamzerp-1)}{z_0-z_1}
\\
& +\frac{\lamzer(\lamtwop-1)+\lamtwo(\lamzerp-1)}{z_0-z_2},
\end{split}
\label{transq}
\end{align}
as well as the transformed scheme
\begin{equation}
\left[\begin{array}{ccccl}
1                & 1                & 1                & 1                                     &     \\
z_0	         & z_1              & z_2	       &\infty                                 & ; z \\
0                & 0                & 0                &\laminf + \lamzer + \lamone + \lamtwo  & ; q \\
\lamzerp-\lamzer & \lamonep-\lamone & \lamtwop-\lamtwo &\laminfp + \lamzer + \lamone + \lamtwo &
\end{array}\right].
\end{equation}
With the assumptions $z_0 = 0$ and $z_1 = 1$, this 
scheme corresponds to the Heun equation 
$G'' + P(z) G' + Q(z)G = 0$ in normal form, 
where
\begin{align}
\begin{split}
P(z) & = \frac{c}{z}
     + \frac{d}{z - 1}
     + \frac{1+a+b-c-d}{z - z_2} \\
Q(z) & =  \frac{ab}{(z-1)(z-z_2)}
         - \frac{q z_2}{z(z-1)(z-z_2)}.
\end{split}
\label{formofRS2}
\end{align}
Here the transformed scheme
\begin{equation}
\left[\begin{array}{ccccl}
1         & 1         & 1         & 1             &     \\
0         & 1         & z_2	  &\infty         & ; z \\
0         & 0         & 0         & a             & ; q \\
1-c	  & 1-d       & c+d-a-b   & b &
\end{array}\right]
\end{equation}
is expressed in terms of the constants $a$, $b$, $c$, and $d$ 
which may be related to the above exponent pairs 
$\{\lambda_k,\lambda_k'\}$. The normal form of 
(\ref{eq:notimeZerilli}) then has the scheme
\begin{equation}
\left[\begin{array}{ccccl}
1  & 1   & 1        & 1          &            \\
0  & 1   & -3/(2n)  &\infty      & ; \rho     \\
0  & 0   & 0        & 2 -\ell    & ; 1 - 4n/3 \\
0  & 0   & -3       & \ell + 3   &
\end{array}\right].
\end{equation}

While the preceding analysis both addresses the structure of the 
time-independent Zerilli equation and reveals the asymptotic 
behavior of the solutions near any given singular point, it does 
not provide concrete analytical expressions for the solutions 
$v^\mathrm{polar}_{L,R}$ considered in the main text. To obtain
such expressions, we use the intertwining operators 
\cite{AndersonPrice1991}
\begin{equation}
D_\pm = \left(1-\frac{1}{\rho}\right)\frac{d}{d\rho} 
    \pm \left[\frac{2}{3} n (n+1) + \frac{3(\rho-1)}{\rho^2
        (3 + 2 n \rho)}\right].
\end{equation}
Using our earlier solutions $v^\mathrm{axial}_{L,R}(\rho)$ to 
the time-independent Regge-Wheeler equation, we then get 
corresponding solutions $v^\mathrm{polar}_{L,R}(\rho) 
\equiv D_+ v^\mathrm{axial}_{L,R}(\rho)$ to 
(\ref{eq:notimeZerilli}) by direct application of $D_+$ and 
the identity
\begin{equation}
\frac{d}{dz}\; {}_2F_1(a,b;c;z) = \frac{ab}{c}\; {}_2F_1(a+1,b+1;c+1;z).
\end{equation}
Therefore, we have also expressed the relevant polar solutions in 
terms of the Gauss--hypergeometric function ${}_2 F_{1}$. The 
analysis above then shows that we are likewise able to express 
solutions to a particular instance of the Heun equation in terms 
of hypergeometric functions. 

%%%%%%%%%%%%%%%%%%%%%%%%%%%%%%%%%%%%%%%%%%%%%%%%%%%%%%%%%%%%%%%
%
%  REFERENCES CITED
%
%%%%%%%%%%%%%%%%%%%%%%%%%%%%%%%%%%%%%%%%%%%%%%%%%%%%%%%%%%%%%%%

\bibliographystyle{plain}
\end{document}